\documentclass[
reprint,
bibnotes,
amsmath,amssymb,
aps,
prd
]{revtex4-2}

\usepackage{graphicx}
\usepackage{dcolumn}
\usepackage{enumitem}
\usepackage{bm}
\usepackage[many]{tcolorbox}
\usepackage{shuffle}
\usepackage{xcolor}
\usepackage{physics}
\usepackage{tabularx}
\usepackage{diagbox}
\usepackage{subcaption}
\usepackage{multirow}
\usepackage{caption}
\usepackage{tikz}
\usetikzlibrary{decorations.markings}
\usetikzlibrary{decorations.pathmorphing}
\usetikzlibrary{intersections}
\usetikzlibrary{calc}
\usepackage{verbatim}

\usepackage{CJK}
\definecolor{Maroon}{RGB}{128, 0, 0}
\definecolor{Green}{rgb}{0.0, 0.5, 0.0}
\definecolor{Purple}{rgb}{0.5, 0.0, 0.5}
\definecolor{Blue}{rgb}{0.0, 0.0, 0.6}

\begin{document}

\begin{CJK*}{UTF8}{}
\CJKfamily{gbsn}

\title{Correlator Polytopes}

\author{Carolina Figueiredo$^{1}$}
\email{cfigueiredo@princeton.edu}
\author{Francisco Vaz\~ao$^{2}$}
\email{fvvazao@mpp.mpg.de}

\affiliation{$^{1}$Jadwin Hall, Princeton University, Princeton, NJ, 08540, USA \\
$^{2}$Max-Planck-Institut f\"ur Physik, Werner-Heisenberg-Institut, Boltzmannstr, 8, 85748 Garching, Germany}

\begin{abstract}
Recently, ``cosmohedra" have been introduced as polytopes underlying the cosmological wavefunction for conformally coupled Tr($\Phi^3$) theory in FRW cosmologies, generalizing associahedra for flat space scattering amplitudes. In this letter we show that correlation functions are also directly captured by a new polytope -- the \textit{Correlatron}. The combinatorics of correlation functions is an interesting blend of flat space scattering amplitudes and wavefunctions. This is reflected in the correlatron geometry, which is a one-higher dimensional polytope sandwiched between cosmohedron and associahedron facets. We provide an explicit embedding for the correlatron, which is a natural extension of the ``shaving'' picture for cosmohedra to one higher dimension.  As a byproduct, we also define ``graph correlahedra'' as polytopes for the contribution to correlators from any fixed graph. We show how the canonical form of these polytopes directly computes the graph correlator, without the power of two weights seen in previous geometric formulations. Finally, we give a prescription for extracting the full correlator from the canonical form of the correlatron.
\end{abstract}
\maketitle
\end{CJK*}

\noindent{\bf Introduction.}--- In the context of inflationary cosmology, observable predictions can be formulated in terms of spatial correlations of fields. According to the Born rule,  an $n$-point correlation function is defined by integrating $n$ field insertions at a given time surface, against the mod squared of the vacuum wavefunction,  $\Psi[\Phi]$, computed via the Feynman path integral as:
\begin{equation}
    \Psi = \exp\left\{ \sum_{n\geq 2} \int \prod_{i=1}^n\, d^d k_i \, \Psi_n[\vec{k}_{i} ]\, \delta^d\left(\textstyle{\sum_i} \vec{k}_i\right)  \right\},
    \label{eq:wavefunc}
\end{equation}
where $\Psi_n$ is the wavefunction coefficient for $n$ field insertions at the boundary. In this paper, we focus on a theory of massless scalars interacting via a cubic interaction --  usually called Tr$(\Phi^3)$ theory \cite{ABHY} -- in an Minkowski background. As shown in \cite{CosmoPolys}, upon a field redefinition the action for this theory is conformally equivalent to that of a conformally coupled scalar in FLRW spacetime (as shown in \cite{CosmoPolys}). The resulting action is given by: 
\begin{equation}
    \mathcal{S}[\phi]=\int d^d x d\eta \, \frac{1}{2}\text{Tr}\left( \partial \phi\right)^2 - \frac{\lambda_3(\eta)}{3}\text{Tr}\, \phi^3 \, ,
    \label{eq:action}
\end{equation}
where for flat-space the coupling $\lambda(\eta) \equiv \lambda$ is a constant; we further comment on the differences with FLRW theories in appendix A. The study of this toy model, with general polynomial interactions, has gained a lot of attention in recent times \cite{CosmoPolys,Cosmohedron,Anninos:2014lwa,CosmoBoot,DiffEq_CosmoCorr,KinFlow,GeometryCosmoCorr,Benincasa:2024lxe,Goodhew:2020hob,CosmoReview,KinFlowLoop,IRSub,MelvillePajer,PajerUnitarityLocality,GoodhewCutCosmoCorr,VicenteVazao:2025cwb,Benincasa:2025uie,LoopInts,PokrakaDEs,CosmoSMatrix,Sachs2,CosmoTreeTheo,Meltzer,Bittermann,CespedesScott,Sachs3,HiddenZerosWF,De:2024zic,Baumann:2025qjx,Glew:2025otn,Glew:2025ugf,Benincasa:2019vqr,Fan:2024iek}.
The wavefunction coefficients $\Psi_n[\vec{k}_{i}]$ can be schematically represented as sums of all cubic Feynman diagrams anchored at the $n$-points on the boundary. The singularities of $\Psi_n[\vec{k}_{i} ]$ are given by the locus where the sum of the $|\vec{k}_{i}|$ entering a given sub-graph add to zero~\footnote{Of course, these are outside the physical kinematic locus where all the $|\vec{k}_{i}|>0$.}. Graphically, we can represent these singularities by tubes enclosing the respective sub-graphs such that a given term in $\Psi_n[\vec{k}_{i} ]$ has singularities corresponding to maximal collections of non-overlapping subgraphs on a given graph. These are called tubings of a graph, where two tubes are non-overlapping if one is fully inside the other, or completely disjoint. Therefore the contribution to the integrand of a given graph is simply the sum over all tubings of the graph. As explained in \cite{Cosmohedron}, we can recast this definition of $\Psi_n[\vec{k}_{i} ]$ by thinking of diagrams as triangulations of the momentum polygon, $P_n$, obtained by consecutively drawing the $\vec{k}_i$ tip-to-toe. In this picture, the tubings are recognized as maximal collections of nested (non-overlapping) subpolygons inside $P_n$ -- these collections are also referred to as Russian-dolls \footnote{At loop-level the momentum polygon can be replaced by a surface, $\mathcal{S}_n$, which specify the order in the topological expansion under consideration, and everything else holds the same -- diagrams correspond to triangulations of $\mathcal{S}_n$ and russian dolls are given by maximal collections of nested (non-overlapping) subsurfaces of $\mathcal{S}_n$. For example, at one-loop $n$-points $\mathcal{S}_n$ would be the one-punctured disk with $n$-marked points on the boundary.}. This definition of $\Psi_n[\vec{k}_{i} ]$ reproduces the \textit{old-fashioned perturbation theory} representation of the wavefunction given in \cite{CosmoPolys}. A given Russian doll, $\mathcal{R}$, produces singularities corresponding to the perimeters of the sub-polygons in $\mathcal{R}$, and the full integrand is given by the sum over all possible Russian dolls. 

In \cite{Cosmohedron}, it was shown that the combinatorics of Russian dolls are nicely captured by a polytope -- the cosmohedron. In this paper, we describe a new geometry, closely related to the cosmohedron, capturing the combinatorics of all the terms contributing to the \textit{correlator}, which is the direct physical observable.

Applying the Born rule to \eqref{eq:wavefunc}, we can write the correlator as (see App. \ref{sec:AppCorrelator})
\begin{equation}
 \langle \Phi_1 \Phi_2 \cdots \Phi_n \rangle = 2\Psi_n + \sum_{\mathcal{C} \neq  \emptyset} \prod_{(i,j)\in\mathcal{C}} \frac{1}{2k_{i,j}} \prod_{P \in \mathcal{P}_\mathcal{C}} 2\Psi_P,
    \label{eq:Corr}
\end{equation}
where we have the respective $n$-point wavefunction, $\Psi_n$, together with a sum over all possible collections of non-overlapping chords $\mathcal{C}$ -- partial and full triangulations -- which divide the original $n$-gon into a collection of subpolygons, $\mathcal{P}_\mathcal{C}$, from which we get the product of the respective lower-point wavefunctions, $\Psi_{P}$, for all $P \in \mathcal{P}_\mathcal{C}$; as well as a factor of $1/k_{i,j}$ for each chord $(i,j) \in \mathcal{C}$, with $k_{i,j} = |\vec{k}_i +\vec{k}_{i+1} + \cdots +\vec{k}_{j-1}|$. From \eqref{eq:Corr}, it is clear that the singularities of the correlator are associated with \textit{tubes} as well as \textit{chords}. 
\begin{figure*}[t]
    \centering
    \includegraphics[width=\linewidth]{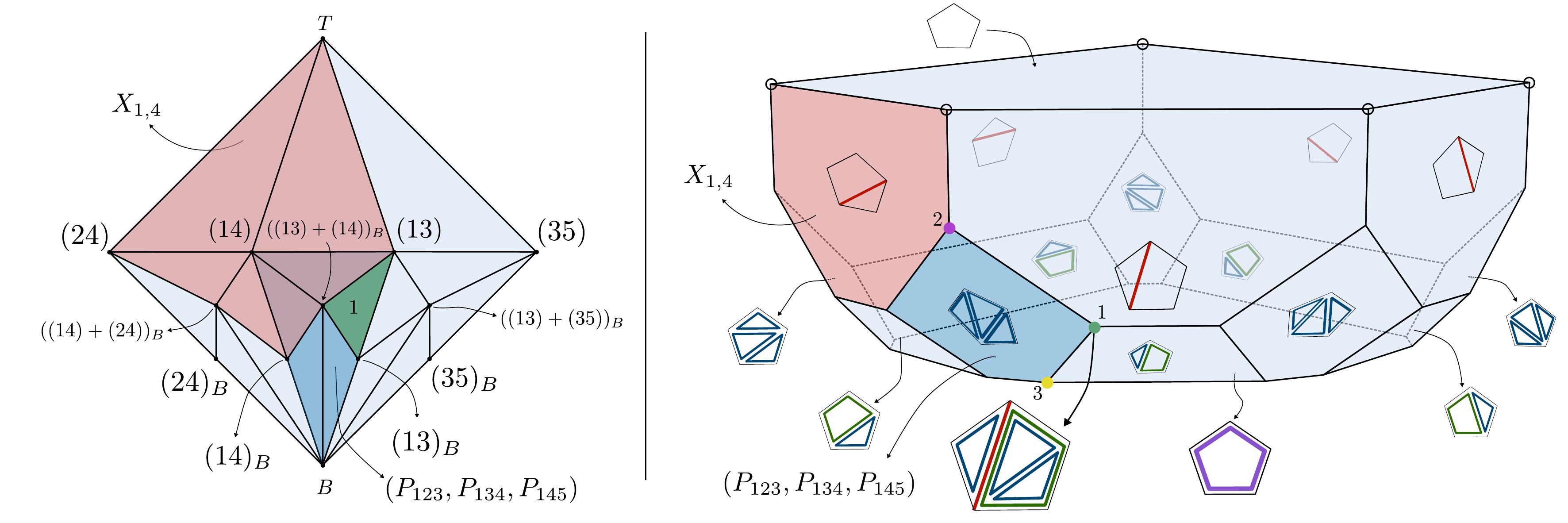}
    \caption{\textbf{(Left)} $2$-dimensional projection of the $3$-dimensional fan of the $5$-point  correlatron. On top of the associahedron rays, $(i,j)$, and the cosmohedron rays, $(i,j)_B$, we have the bottom and top rays, $B$ and $T$. \textbf{(Right)} Realization of the (truncated) $5$-point correlatron, with the appropriate facet labellings in terms of sub-polygon tilings or single chords. In red we highlight the facet for chord $(1,4)$, with facet inequality $X_{1,4} \geq 0$; and in blue the facet corresponding to the triangle-tiling given by triangulation $\{(1,3),(1,4)\}$, labelled by $(P_{123},P_{134},P_{145})$, and with facet inequality $X_{1,3} + X_{1,4} + X_B\geq \epsilon_{\{(1,3),(1,4)\}}$. 
    We highlight three vertices ($1, 2$ and $3$) on the right, and provide the explicit computation of their contribution to the correlator in appendix \ref{sec:AppExamples}.}
    \label{fig:5pt}
\end{figure*}

In \cite{Cosmohedron}, it was proposed that the combinatorics of the different terms in the correlator is nicely captured by a polytope which is a \textit{sandwich} between an associahedron and a cosmohedron. In this letter, we give a precise definition for this polytope -- which we call the \textit{correlatron}. We present explicit inequalities for cutting out this polytope at all $n$, both at tree-level and one-loop. As with cosmohedra, we expect both the definition and the embedding to generalize to all loop orders built on the existence of surfacehedra~\cite{Surfacehedra}. We similarly define \textit{graph correlahedra}, corresponding to the polytopes which encode a single graph's contribution. Quite remarkably, the correlatron embedding for graph correlahedra allows us to compute the graph correlator directly as a flat-out canonical form -- without the need for any weights as it was previously propposed in \cite{GeometryCosmoCorr}. We explain this construction in detail in appendix \ref{sec:PhysKinGraphCorr}. 
\\ \\
\noindent{\bf Combinatorial definition of the Correlatron.}---Let's recall the familiar polytopes we have for the amplitude and the wavefunction -- the associahedron~\cite{Arkani-Hamed:2017mur} and the cosmohedron \cite{Cosmohedron}, respectively. Associahedra are polytopes whose faces are labeled by collections of non-overlapping chords, $\mathcal{C}$, following the rule that $\mathcal{C}^{\prime}$ is a face of $\mathcal{C}$, if $\mathcal{C} \subset \mathcal{C}^{\prime}$; which in particular means that codim-1 facets are labeled by single chords. On the other hand, cosmohedra are polytopes whose faces are labeled by collections of non-overlapping (but possibly nested) sub-polygons, $\mathcal{P}$, and similarly a collection $\mathcal{P}^{\prime}$ is a face of $\mathcal{P}$, if $\mathcal{P} \subset \mathcal{P}^{\prime}$; in which case the codim-1 facets are labeled by partial triangulations of $P_n$ (with any number of chords) which divide the $n$-gon into disjoint (not nested) subpolygons \cite{Cosmohedron} -- which we here call \textit{subpolygon tilings}. 
The polytope describing the correlator as described in \cite{Cosmohedron}, corresponds to the simplest merging of the combinatorics above: it's a polytope whose faces are labeled by collections, $\{\mathcal{C},\mathcal{P}\}$, of non-overlapping chords, $C$, and subpolygons $P$, where the subpolygons must be compatible with the chords $C$. We have that the correlator polytope satisfies:
\begin{equation}
    \{\mathcal{C}^\prime, \mathcal{P}^\prime\} \text{ is a face of }\{\mathcal{C},\mathcal{P}\} \text{ if } \mathcal{C} \subset \mathcal{C}^\prime \text{ and } \mathcal{P} \subset \mathcal{P}^\prime.
    \label{eq:def_correlatron}
\end{equation} 

In particular, the facets of the correlatron have the same labels as facets of associahedra and cosmohedra -- they are either labeled by single chords in $P_n$, or by collections of subpolygons that tile  $P_n$ (see figure \ref{fig:5pt} for $n=5$ example). In addition, the correlatron has two extra facets: the bottom facet which is a Cosmo$_n$, which is labeled by a subpolygon tiling with a single subpolygon equal to the full $P_n$; and a top facet which is an Assoc$_n$, labelled by an empty collection of chords. Strictly speaking, the associahedron top facet should be thought to be at infinity since the vertices living in this facet do not correspond to terms in the correlator, but for simplicity, we will always draw the truncated version of the polytope, in which the associahedron facet appears. On the other hand, the cosmohedron bottom facet, labeled by $\{P_n\}$, is precisely encoding all the terms entering in the correlator coming from $\Psi_n$, as given in \eqref{eq:Corr} -- $i.e.$ all terms containing the $E_t = |\sum_{i=1}^n \vec{k}_i|$ singularity. Note that from \eqref{eq:def_correlatron}, the only facets compatible with the Cosmo$_n$ facet are those labeled by subpolygon tilings where the collection of chords $\mathcal{C}$ is empty, and the only facets compatible with the Assoc$_n$ top facet have an empty collection $\mathcal{P}$.

Just like for cosmohedra, the facets labeled by tilings of $P_n$ containing \textit{only} triangles are simple polytopes whose vertices label the contributions to the correlator from the graph dual to the triangulation $\mathcal{T}$, defined by the respective collection of triangles. We call these geometries \textit{graph correlahedra}. 
\\ \\
\noindent{\bf Facet Structure.}--- Let us start by considering a facet labeled by a single chord, $c$, which divides $P_n$ into two polygons, $P_c^L$ and $P_c^R$. This facet is then simply the direct product of the correlatrons for each sub-polygon:
\begin{equation}
    {\rm Facet}_c[{\rm Corr}_n] = {\rm Corr}_{P_c^L} \times {\rm Corr}_{P_c^R}\, .
\end{equation}
On the other hand, a facet labelled by a subpolygon tiling of $P_n$, $\mathcal{P}_C$, is the direct product of the cosmohedra associated with each of the sub-polygons $P \in \mathcal{P}_C$, and the graph correlahedra, $\text{GraphCorr}_{\mathcal{G}}$, of the graph $\mathcal{G}$ which is dual to the respective subpolygon tiling. Explicitly,
\begin{equation}
    {\rm Facet}_{\mathcal{P}_C}[{\rm Corr}_n] = \text{GraphCorr}_{\mathcal{G}} \times \prod_{P \in \mathcal{P}_C} {\rm Cosmo}_{P} \, ,
\end{equation}
which makes it clear that the facet where $\mathcal{P}_C = \{P_n\}$, is simply the $n$-point cosmohedron.
We present the full $F$-vector of the correlatron up to $n=8$ in table \ref{tab:FVector}, where the top associahedron facet has been sent to infinity.  
\begin{table}[t]
\centering
\begin{tabular}{c|c|c|c|c|c|c}
 \diagbox{$n$ pts}{Cdim} & 1   & 2     & 3     & 4     & 5     & 6     \\ \hline
4 & 5   & 4     & -     & -     & -     & -     \\
5 & 16  &  40   & 25    & -     & -     & -     \\
6 & 54  & 272   & 419   & 200   & -     & -     \\
7 & 211 & 1750  & 4676  & 5026  & 1890  & -     \\
8 & 923 & 11278 & 45232 & 80366 & 65738 & 20248
\end{tabular}
\caption{\label{tab:FVector}$F$-vectors up to $n=8$.}
\end{table}
\\ \\
\noindent{\bf Correlatron Embeddings.}--- We now show how to cut out the correlatron with inequalities at all $n$. The example for the embedding of the five-points correlatron (see also figure \ref{fig:5pt}) is explained explicitly in appendix \ref{sec:AppExamples}. These embeddings are closely related to the ABHY realization of associahedra \cite{ABHY} and the embeddings of cosmohedra \cite{Cosmohedron}, so let us start by briefly reviewing these constructions. We work at tree-level here and explicitly describe the one-loop case in appendix \ref{app:one-loop}. We expect the all loop generalization to be straightforward, replacing the associahedra with the surfacehedra defined in \cite{Surfacehedra}. 

As proposed in \cite{ABHY}, to each facet of Assoc$_n$, labeled by a chord $(i,j)$, we associated a variable $X_{i,j}$. The dual fan of Assoc$_n$ is then formed by a set of $\vec{g}$-vectors, $\vec{g}_{i,j}$, one per internal chord, and the top-dimensional cones are spanned by collections of $(n-3)$ $\vec{g}$-vectors that are in one-to-one correspondence with triangulations of $P_n$. To embed the polytope in $X_{i,j}$ space we first impose $X_{i,j} \geq 0$, for all $X_{i,j}$, and then intersect this space with the ABHY plane \cite{ABHY}, which is defined as follows: pick a triangulation of $P_n$, $\mathcal{T}^\star$, take all chords $(k,m)$ \textit{not} in $\mathcal{T}^\star$, and consider the plane obtained by taking all $c_{k,m}$, with
\begin{equation}
    c_{k,m} = X_{k,m} + X_{k+1,m+1} -X_{k,m+1}-X_{k+1,m}, 
\label{eq:ABHYplane}
\end{equation}
to be positive constants. This produces an $(n-3)$-dimensional geometry which is the ABHY associahedron. 

In turn, the cosmohedron and its embedding can be simply thought of as a ``shaving" of an underlying associahedron. This shaving can be naturally phrased as a refinement of the associahedron fan which ``blows-up" the different codimension faces into facets. Concretely, starting with a given cone in the fan of Assoc$_n$, we consider a refinement containing all the Assoc$_n$ $\vec{g}$-vectors as well as all possible vectors we get by summing the Assoc$_n$ $\vec{g}$-vectors in the cone (see \cite{Cosmohedron}). Each new ray, $\vec{g}_{\mathcal{C}}$, is then associated with a collection of non-overlapping chords, $\mathcal{C}$, and is given in terms of the original $\vec{g}$-vectors as $\vec{g}_{\mathcal{C}} = \sum_{(i,j) \in \mathcal{C}} \vec{g}_{i,j}$. As mentioned earlier, any such collection $\mathcal{C}$ specifies a subpolygon tilling of $P_n$, which is the label we associate to the facet normal to $\vec{g}_\mathcal{C}$. 

With this prescription, a facet of Cosmo$_n$, associated with a collection $\mathcal{C}$, comes with the facet inequality:
\begin{equation}
    \sum_{(i,j) \in \mathcal{C}} X_{i,j} \geq \epsilon_\mathcal{C}\, ,
    \label{eq:cosmo_facets}
\end{equation}
where the $\epsilon_\mathcal{C}$ are positive constants, such that $\epsilon_\mathcal{C} \ll c_{k,m}$. In addition, given any two collections $\mathcal{C}$ and $\mathcal{C}^\prime$, such that $\mathcal{C} \nsubseteq \mathcal{C}^\prime$ (or vice-versa), the respective $\epsilon_\mathcal{C}$ and $\epsilon_{\mathcal{C}^\prime}$ obey
\begin{equation}   \epsilon_\mathcal{C}+\epsilon_{\mathcal{C}^{\prime}} < \epsilon_{\mathcal{C}\cup \mathcal{C}^{\prime}} + \epsilon_{\mathcal{C}\cap \mathcal{C}^{\prime}}\, ,
    \label{eq:cosmo_eps_ineqs}
\end{equation}
for all collections $\mathcal{C}$ and $\mathcal{C}^\prime$ except for the cases in which $\mathcal{C}$ is completely on one side of $C \cap C^{\prime}$, and $C^{\prime}$ is completely on the other, where instead the above inequality turns into an equality of the same form:
\begin{equation}   \epsilon_\mathcal{C}+\epsilon_{\mathcal{C}^{\prime}} = \epsilon_{\mathcal{C}\cup \mathcal{C}^{\prime}} + \epsilon_{\mathcal{C}\cap \mathcal{C}^{\prime}}\, ,
    \label{eq:cosmo_eps_eqs}
\end{equation}
The presence of equalities is what leads to the non-simple vertices of the cosmohedron.

Now we turn to the embedding of the correlatron. From the combinatorial definition, it is clear that this object has both types of facets, the ones for individual chords, $X_{i,j}$, as well as for subpolygon tilings labeled by collections $\mathcal{C}$, from \eqref{eq:facet_ineqs}. In addition, it also has the facet associated to the $\{P_n\}$ tiling, which we label by $X_B$, and, if considering the truncated version, the associahedron facet labeled by an empty collection of chords, $X_T$. To facet $X_B$, we associate a new $\vec{g}$-vector, $\vec{g}_B$, which is independent from all the $g_{i,j}$'s; as for $X_T$ we have $\vec{g}_T = -\vec{g}_B$, as the associahedron facet is truncating the geometry in this extra dimension. Thus, for a given $n$, the correlatron lives in one higher dimension than the Cosmo$_n$/Assoc$_n$. 

We can then define the correlatron fan by starting with the associahedron fan living in one extra dimension such that $\vec{g}_B = -\vec{g}_T$ are orthogonal to all vectors $\vec{g}_{i,j}$. Then we consider the cosmohedron refinement of the associahedron fan and add it to this picture by adding to each cosmo $\vec{g}$-vector, $\vec{g}_B$. This is for all collections of non-overlapping chords, $\mathcal{C}$, we associated the vector $\vec{g}_\mathcal{C}^B = \sum_{(i,j) \in \mathcal{C}} \vec{g}_{i,j} +\vec{g}_B $. An example of the fan for the $5$-point correlatron is shown on the left of figure \ref{fig:5pt}.   
\begin{figure*}[t]
    \centering
    \includegraphics[width=\linewidth]{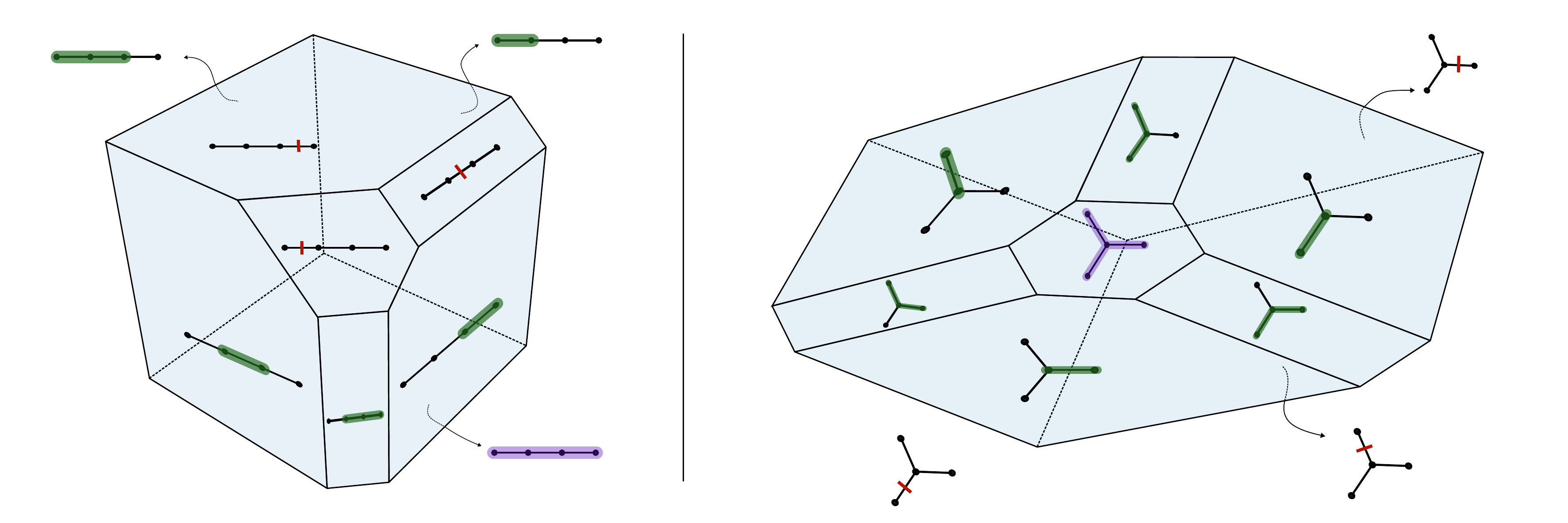}
    \caption{\textbf{(Left)} Graph correlahedra at $6$-points for triangulations of the hexagon whose dual graph is a chain.  \textbf{(Right)} Graph correlahedra at $6$-points for triangulation $\{(1,3),(3,5),(1,5)\}$ (and cyclic). In both cases, the facets are associated with internal edges (red) or tubings enclosing more than a single site (green/purple).}
    \label{fig:graphCorrs}
\end{figure*}

From this fan, we can read off the facet inequalities
\begin{equation}
    X_{i,j} \geq 0 , \,\, \sum_{(a,b) \in \mathcal{C}} X_{a,b} +X_B \geq \epsilon_\mathcal{C} , \,\, X_B\geq \epsilon_B, \,\, X_T \geq 0, 
\label{eq:facet_ineqs}
\end{equation}
for all chords $(i,j)$, and compatible collections $\mathcal{C}$. Here again, we consider the $X_{i,j}$'s to be in the ABHY plane and we take $\epsilon_\mathcal{C} \ll c_{k,m}$. In addition, we have $X_B + X_T = H$, and we take the facet $X_T$ to infinity by taking $H\to \infty$. 

To correctly reproduce the combinatorics of the correlator, the constants $\epsilon_\mathcal{C}$ have to satisfy an interesting set of inequalities. First of all, they need to satisfy the inequalities and equalities of the cosmohedron \eqref{eq:cosmo_eps_ineqs} -- which directly imply that the correlatron is \textit{not} a simple polytope. Furthermore, by allowing vertices where individual chord facets meet those of subpolygon tillings, we find that the $\epsilon_\mathcal{C}$ must further satisfy:
\begin{equation}
    \epsilon_\mathcal{C} > \epsilon_{\mathcal{C}^{\prime}} \, ,
    \label{eq:new_eps_ineqs}
\end{equation}
where $\mathcal{C}^{\prime} \subset \mathcal{C}$, and therefore is not in contradiction with \eqref{eq:cosmo_eps_ineqs}. From \cite{Cosmohedron}, we know that the $\epsilon_\mathcal{C}$ equalities, coming from \eqref{eq:cosmo_eps_ineqs}, are automatically satisfied when we write:
\begin{equation}
    \epsilon_\mathcal{C} = \sum_{P \subset \mathcal{C}} \delta_P \, ,
    \label{eq:eps_to_delta_map}
\end{equation}
where $P$ are the sub-polygons in the partial triangulation defined by $\mathcal{C}$. In turn, the inequalities \eqref{eq:cosmo_eps_ineqs} and the new inequalities \eqref{eq:new_eps_ineqs} become, respectively:
\begin{equation}
    \delta_P+\delta_{P^{\prime}} < \delta_{P\cup P^{\prime}} + \delta_{P\cap P^{\prime}}\, , \quad \delta_P+\delta_{P^{\prime}} > \delta_{P\cup P^{\prime}} \, ,
    \label{eq:delta_ineqs}
\end{equation}
where we must set $\delta_{P_n}=\epsilon_B$. A parameterization of the $\delta$'s  obeying the cosmohedron condition \eqref{eq:delta_ineqs}, is \cite{Cosmohedron}
\begin{equation}
    \delta_P = \delta (n - \# P)^2\, ,
    \label{eq:delta_param}
\end{equation}
where $\# P $ is the number of sides of $P$. This also trivially satisfies the new constraints needed for the correlatron, since sub-polygons with a lower number of sides have higher value.
\\ \\
\noindent{\bf Graph Correlahedra.}--- Let us now discuss the geometry describing the contributions to the correlator from a single graph -- the \textit{graph correlahedron}.

All the terms contributing to the correlator from a particular triangulation, $\mathcal{T}$, share a set of singularities: the perimeters of the triangles $t \subset \mathcal{T}$. Graph correlahedra are then geometries whose facets label the remaining singularities entering the correlator for a fixed graph, which are either single chords in $\mathcal{T}$, or non-triangle sub-polygons compatible with $\mathcal{T}$ -- $i.e.$ whose edges are either edges of $P_n$ or chords in $\mathcal{T}$ -- corresponding to tubes of the graph that enclose more than a single site. 

Combinatorially, the face structure of graph correlahedra follows \eqref{eq:def_correlatron}, where now $\mathcal{C}$ is restricted to the chords in the triangulation, $\mathcal{T}$, and $\mathcal{P}$ to a set of non-triangle subpolygons compatible with $\mathcal{C}$.

Since this geometry corresponds to a facet of the correlatron -- labeled by the collection of triangles dual to it -- we can directly read off its embedding via the correlatron inequalities. In particular, localizing on a facet associated with a triangle-tiling, specified by some triangulation $\mathcal{T}$, corresponds to 
\begin{equation}
    \sum_{(i,j) \in \mathcal{T}} X_{i,j} + X_B = \sum_{t \subset \mathcal{T}} \delta_t \, ,
    \label{eq:graphCorrEq}
\end{equation}
where $t$ is labelling the triangles in $\mathcal{T}$. Now the facets labelled by subpolygon tilings that intersect \eqref{eq:graphCorrEq} are those associated with partial triangulations coarsening $\mathcal{T}$. Any such partial triangulation, $T^\prime$, in general contains triangles in $\mathcal{T}$ and non-triangle subpolygons, $P$, such that we can write the respective facet inequality as
\begin{equation}
     \sum_{(i,j) \in T^{\prime}} X_{i,j} +X_B \geq \sum_{t \subset T^{\prime}} \delta_t + \sum_{P \subset T^{\prime}} \delta_P \, .
     \label{eq:genFacetIneq}
\end{equation}
However, just like for cosmohedra, one can easily check that all the inequalities \eqref{eq:genFacetIneq} for partial triangulations that contain \textit{more than one} non-triangle subpolygon $P$, are \textit{redundant}, under the support of \eqref{eq:graphCorrEq}. Therefore the only facet inequalities of the graph correlahedron correspond to those with a \textit{single} non-triangle subpolygon, $P$, which are therefore associated with the singularity of the associated tube, $i.e.$ the perimeter of $P$. 

We also have the facets associated with single chords entering in $\mathcal{T}$, with facet inequalities, $X_{i,j} \geq 0$ for all $(i,j) \in \mathcal{T}$; as well as the facet corresponding to the subpolygon tilling $\{P_n\}$ which is associated with the total energy singularity, whose facet inequality is $X_B \geq 0$.

Now to extract the correlator for the graph associated with $\mathcal{T}$, we start by associating the correct singularity to each facet -- to a given chord $(i,j)$ we associate a factor $k_{i,j}$, to a non-triangle subpolygon $P$, we associate its perimeter, $\mathcal{P}_P$, and to the bottom facet we associate $\mathcal{P}_{P_n} = |\sum_{i=1}^n \vec{k}_i|$. Then we have
\begin{equation}
    \langle \Phi_1 \Phi_2 \cdots \Phi_n \rangle_{\mathcal{T}} = \prod_{t\subset\mathcal{T}} \frac{1}{\mathcal{P}_t} \times \Omega\left(\text{GraphCorr}_\mathcal{T}\right),
\end{equation}
where $\Omega$ is the canonical form of the graph correlahedron polytope. Note that here, we have not explicitly imposed the relations between the $\mathcal{P}_P$'s which come from the fact that they are all perimeters of subpolygons of the same underlying momentum $n$-gon. In appendix \ref{sec:PhysKinGraphCorr}, we explain how doing this still allows us to compute the correlator in physical kinematics directly from the canonical form of the resulting geometry. 
\\ \\
\noindent{\bf Extracting the correlator from the correlatron.}--- 
Just like for the cosmohedron, a standard canonical form for the correlatron, which associates a single singularity to each facet, does not correctly compute the correlator. Instead, there is a more interesting recipe which we now explain. 

Firstly, to extract the result, we blow-up all the non-simple vertices, to obtain a simple polytope which we call the permuto-correlatron. This object is obtained by turning all the equalities of the form \eqref{eq:cosmo_eps_eqs} into strict inequalities. 

Just like for the graph correlahedra, for the facets labeled by a single chord, $(i,j)$, we associate a factor: $X_{i,j}^{-1} \equiv k_{i,j}^{-1}$; and for the bottom facet we associate the \textit{total energy} singularity: $
    X_B^{-1} \equiv E_T^{-1} = \mathcal{P}_{P_n}^{-1}$.

As for the facets labeled by subpolygon tilings defined by a collection of chords $\mathcal{C}$, we assign the factor:
\begin{equation}
\frac{1}{X_\mathcal{C}} \equiv \frac{1}{n_C}\left[ \sum_{(P|P^\prime) \subset \mathcal{C}} \frac{1}{{\cal P}_P {\cal P^\prime}_{P^\prime}} + \prod_{P \subset \mathcal{C}}\frac{1}{\mathcal{P}_P}\right],
\label{eq:WF_permuto_sings}
\end{equation}
where $n_C$ is the number of non-triangle sub-polygons in the tiling, except when the tiling contains \textit{only} triangles, in which case $n_C \equiv 1$. The sum is over all pairs of subpolygons, $P$ and $P^\prime$, in $\mathcal{C}$, such that $P$ and $P^\prime$ share an edge. As compared to the factor associated with cosmohedron facets, for the correlatron, we now have an extra term given by the product over the perimeters of \textit{all} subpolygons $P \subset \mathcal{C}$. 

\vspace{2mm}
Computing the canonical form for the permuto-correlatron with the facet factors described above, we find that there is a single piece that has the correct units as well as no double or higher poles -- this is precisely the $n$-point correlator (up to an overall factor of 2):
\begin{equation}
    \langle \Phi_1 \Phi_2 \cdots \Phi_n \rangle =  \Omega\left(\text{PermCorr}_{n}\right) \vert_{\text{single poles}, |k|^{-2n+5}}.
\end{equation}
\\ 
\noindent{\bf Outlook.}--- The discovery of the cosmohedron and the correlatron suggest many obvious avenues for future exploration. Many of the open questions hold for both objects, so we can recapitulate some of the questions asked in \cite{Cosmohedron} about cosmohedra. The definition of these cosmological geometries begins with associahedra as a ``base'', and the final geometries are obtained by adding ``shaving'' inequalities. Is there a more natural definition that does not privilege the base associahedron, and gives a more uniform picture of the inequalities? Perhaps the most striking feature of associahedra is that they are naturally Minkowski sums of simple pieces -- this is the first hint of the existence of u-variables and binary geometries which yield the generalization of particle to string amplitudes. What if anything is the analogue of this set of ideas for the new cosmological objects?

There is an especially fascinating question that is related to the correlator. In all the discoveries connecting positive geometries to scattering amplitudes and wavefunctions, an obvious question has always been: why should such geometries exist for ${\cal A}$ or $\Psi$, when the actual observables, given by the Born rule, are associated with $|{\cal A}|^2$ and $|\Psi|^2$? Is there anything in the world of combinatorics and geometry that asks for the Born rule? With the correlatron, we have a glimpse of an answer to this question. We find it remarkable that from the correlatron embedding, precisely the canonical form for the graph correlatron produces the correct graph correlator, without any factor of two weights that usually arise from the squaring in the Born rule, and that have been necessary for previous formulations of the correlator from geometry~\cite{GeometryCosmoCorr}. For the sum over all graphs, both the cosmohedron and the correlatron call for a deeper understanding of the peculiar rules that have been found for extracting the wavefunction and correlator from the canonical form, and in particular the generalization to ``perimeter kinematics'' plays a crucial role. This might give us a window into combinatorial-geometric origin of the Born rule.
\\ \\
\noindent {\bf Acknowledgements.---}
We thank Nima Arkani-Hamed, Federico Ardilla, Paolo Benincasa, Ross Glew, Austin Joyce, and Hayden Lee for discussions. C.F. is supported by FCT - Fundacao para a Ciencia e Tecnologia, I.P. (2023.01221.BD and DOI  https://doi.org/10.54499/2023.01221.BD). F.V. is supported by the European Union (ERC, UNIVERSE PLUS, 101118787). Views and opinions expressed are however those of the author(s) only and do not necessarily reflect those of the European Union or the European Research Council Executive Agency. Neither the European Union nor the granting authority can be held responsible for them.

\onecolumngrid
\appendix

\section{Correlation functions from wavefunction}
\label{sec:AppCorrelator}
The correlation functions can be computed using the wavefunction in \eqref{eq:wavefunc} as an input to the Born rule:
\begin{equation}
\left\langle\Phi_{\vec{k}_1} \cdots \Phi_{\vec{k}_n}\right\rangle= \mathcal{N}\, \displaystyle\int \mathcal{D} \Phi\, \Phi_{\vec{k}_1} \cdots \Phi_{\vec{k}_n}|\Psi[\Phi]|^2\, ,
\label{eq:BornRule}
\end{equation}
where $\mathcal{N}$ is just the normalization computed with no field insertions. The mod square of the wavefunction can be written as:
\begin{equation}
    |\Psi[\Phi]|^2 = \exp\left\{ \sum_{n\geq 2} \int \prod_{i=1}^n\, d^d k_i \,\Phi_1\cdots \Phi_n\, 2\,{\rm Re}\{\Psi_n[\vec{k}_{i} ]\}\, \delta^d\left(\textstyle{\sum_i} \vec{k}_i\right)  \right\},
    \label{eq:wavefunc_mod_sq}
\end{equation}
where for the case under consideration, since the wavefunction coefficients are real, we have ${\rm Re}\{\Psi_n\}=\Psi_n$. For the case where the background is FLRW, the wavefunction coefficients have an overall complex phase which changes their real part, and thus directly affect their contribution to the correlator \cite{Goodhew:2024eup, Thavanesan:2025kyc}, which we are not discussing in this work. Plugging \eqref{eq:wavefunc_mod_sq} in \eqref{eq:BornRule}, we can easily recognize that the correlator we are computing is just the correlation function in a theory whose action has infinitely many polynomial interaction terms, with the coefficient of $n$-th term being $2\,\Psi_n$, and whose two-point function is then simply $1/(2 \Psi_2)$.

Therefore, we can compute the contributions to the correlator, $ \langle \Phi_1 \Phi_2 \cdots \Phi_n \rangle$, on a graph by graph basis, where for any given graph, $\mathcal{G}_n$, the pure ``contact part'' is simply the contribution of $\mathcal{G}_n$ to the wavefunction coefficient $2\Psi_n$, and the pieces with poles are simply given by products of lower point wavefunction coefficients, which diagrammatically are associated to all possible ways of deleting the edges of $\mathcal{G}_n$. 

Let us look at the simplest example at $4$-points. In this case, we can choose the ``s-channel'' diagram corresponding to the triangulation of the momentum polygon, $P_4$, containing chord $(1,3)$. The dual graph is then a two-site chain with the internal propagator with momentum $k_{1,3} =|\vec{k}_1+\vec{k}_2| =|\vec{k}_3+\vec{k}_4|$, from which we get the following contribution:
\begin{equation}
\begin{aligned}
    \langle\Phi_1 \Phi_2 \Phi_3 \Phi_4\rangle \bigg \vert_{\begin{gathered}
    \begin{tikzpicture}[line width=0.6,scale=0.27,line cap=round,every node/.style={font=\scriptsize}]
        \coordinate (L1) at (-1,0);
        \coordinate (R1) at (1,0);
        \draw (L1) -- (R1);
        \filldraw[color=black,fill=black] (L1) circle[radius=0.2];
        \filldraw[color=black,fill=black] (R1) circle[radius=0.2];
\end{tikzpicture}
\end{gathered}} &= \frac{1}{16 k_1 k_2 k_3 k_4}\left(2\Psi_4(\vec{k}_1,\vec{k}_2,\vec{k}_3,\vec{k}_4) \bigg \vert_{\begin{gathered}
    \begin{tikzpicture}[line width=0.6,scale=0.27,line cap=round,every node/.style={font=\scriptsize}]
        \coordinate (L1) at (-1,0);
        \coordinate (R1) at (1,0);
        \draw (L1) -- (R1);
        \filldraw[color=black,fill=black] (L1) circle[radius=0.2];
        \filldraw[color=black,fill=black] (R1) circle[radius=0.2];
\end{tikzpicture}
\end{gathered}}+ \frac{2\Psi_3(\vec{k}_1,\vec{k}_2,\vec{k}_{1,3})2\Psi_3(\vec{k}_3,\vec{k}_4,\vec{k}_{1,3})}{2k_{1,3}}\bigg \vert_{\begin{gathered}
    \begin{tikzpicture}[line width=0.6,scale=0.27,line cap=round,every node/.style={font=\scriptsize}]
        \coordinate (L1) at (-1,0);
        \coordinate (R1) at (1,0);
        \draw (L1) -- (R1);
        \draw[color=Maroon,thick] (0,-0.2) -- (0,0.2);
        \filldraw[color=black,fill=black] (L1) circle[radius=0.2];
        \filldraw[color=black,fill=black] (R1) circle[radius=0.2];
\end{tikzpicture}
\end{gathered}}\right)\,  \\
&= \frac{1}{16 k_1 k_2 k_3 k_4}\left( \begin{gathered}\begin{tikzpicture}[line width=0.6,scale=0.27,line cap=round,every node/.style={font=\scriptsize},baseline={([yshift=-0.5ex]current bounding box.center)}]
        \coordinate (L1) at (-1,0);
        \coordinate (R1) at (1,0);
        \draw (L1) -- (R1);
        \filldraw[color=black,fill=black] (L1) circle[radius=0.2];
        \draw[color=Green] (L1) circle[radius=0.5];
        \draw[color=Green] (R1) circle[radius=0.5];
        \filldraw[color=black,fill=black] (R1) circle[radius=0.2];
       \draw[thick, Blue]
  ($(L1)!.5!(R1)$) ellipse [x radius=2, y radius=0.8];
\end{tikzpicture}
\end{gathered}  +  \begin{gathered}\begin{tikzpicture}[line width=0.6,scale=0.27,line cap=round,every node/.style={font=\scriptsize},baseline={([yshift=-0.5ex]current bounding box.center)}]
        \coordinate (L1) at (-1,0);
        \coordinate (R1) at (1,0);
        \draw (L1) -- (R1);
        \draw[color=Maroon,thick] (0,-0.2) -- (0,0.2);
        \filldraw[color=black,fill=black] (L1) circle[radius=0.2];
        \draw[color=Green] (L1) circle[radius=0.5];
        \draw[color=Green] (R1) circle[radius=0.5];
        \filldraw[color=black,fill=black] (R1) circle[radius=0.2];
\end{tikzpicture}
\end{gathered} \right) = \frac{1}{16 k_1 k_2 k_3 k_4} \left( \frac{2}{\mathcal{P}_{1,2,3} \mathcal{P}_{1,3,4} \mathcal{P}_{1,2,3,4}} +\frac{2}{k_{1,3}\mathcal{P}_{1,2,3} \mathcal{P}_{1,3,4}} \right)
\end{aligned}
\label{eq:4ptCorrS}
\end{equation}
where the prefactor out-front is simply the product of the $2$-point function for the external lines, which we will in general drop, and focus on the piece between brackets. The perimeter variables, $\mathcal{P}$, entering above are:
\begin{equation}
    \mathcal{P}_{1,2,3} = k_1 + k_2 +k_I, \quad \mathcal{P}_{1,3,4} = k_I + k_3 +k_4, \quad  \mathcal{P}_{1,2,3,4} = E_t = k_1 +k_2 +k_3 +k_4.
\end{equation}

So from \eqref{eq:4ptCorrS} we obtain the contribution to $\langle\Phi_1 \Phi_2 \Phi_3 \Phi_4\rangle$ coming from the $s$-channel diagram, but to compute the full correlator we need to add the contribution from the $t$-channel. This diagram is dual to the triangulation of $P_4$ containing chord $(2,4)$, which then corresponds to the two-site chain with internal propagator $k_{2,4}=|\vec{k}_2+\vec{k}_3|$, and the respective contribution to $\langle\Phi_1 \Phi_2 \Phi_3 \Phi_4\rangle$ can be obtained by a cyclic transformation of \eqref{eq:4ptCorrS}. 

At five-points we have $5$ different triangulations all related by cyclic transformations, and for which the dual graphs are three-site chains. In particular, let us look at the case in which the internal propagators are $k_{1,3} =|\vec{k}_1+\vec{k}_2|$ and $k_{1,4}  =|\vec{k}_1+\vec{k}_2++\vec{k}_3|$. The contribution to the correlator from this graph is then:
\begin{equation}
\begin{split}
       \langle\Phi_1 \Phi_2 \cdots \Phi_5\rangle \bigg \vert_{\begin{gathered}
    \begin{tikzpicture}[line width=0.6,scale=0.27,line cap=round,every node/.style={font=\scriptsize}]
        \coordinate (L1) at (-1,0);
        \coordinate (M) at (1,0);
        \coordinate (R1) at (3,0);
        \draw (L1) -- (R1);
        \filldraw[color=black,fill=black] (L1) circle[radius=0.2];
        \filldraw[color=black,fill=black] (R1) circle[radius=0.2];
        \filldraw[color=black,fill=black] (M) circle[radius=0.2];
\end{tikzpicture}
\end{gathered}} =& \frac{1}{\prod_{i=1}^52k_i }\left(2\Psi_5(\vec{k}_1,\vec{k}_2,\vec{k}_3,\vec{k}_4,\vec{k}_5)\bigg \vert_{\begin{gathered}
    \begin{tikzpicture}[line width=0.6,scale=0.27,line cap=round,every node/.style={font=\scriptsize}]
        \coordinate (L1) at (-1,0);
        \coordinate (M) at (1,0);
        \coordinate (R1) at (3,0);
        \draw (L1) -- (R1);
        \filldraw[color=black,fill=black] (L1) circle[radius=0.2];
        \filldraw[color=black,fill=black] (R1) circle[radius=0.2];
        \filldraw[color=black,fill=black] (M) circle[radius=0.2];
\end{tikzpicture}
\end{gathered}}+ \frac{2\Psi_4(\vec{k}_1,\vec{k}_2,\vec{k}_3,\vec{k}_{1,4})2\Psi_3(\vec{k}_4,\vec{k}_5,\vec{k}_{1,4})}{2k_{1,4}}\bigg \vert_{\begin{gathered}
    \begin{tikzpicture}[line width=0.6,scale=0.27,line cap=round,every node/.style={font=\scriptsize}]
        \coordinate (L1) at (-1,0);
        \coordinate (M) at (1,0);
        \coordinate (R1) at (3,0);
        \draw (L1) -- (R1);
        \filldraw[color=black,fill=black] (L1) circle[radius=0.2];
        \filldraw[color=black,fill=black] (R1) circle[radius=0.2];
        \filldraw[color=black,fill=black] (M) circle[radius=0.2];
        \draw[color=Maroon,thick] (2,-0.2) -- (2,0.2);
\end{tikzpicture}
\end{gathered}} \right.  \\
&\quad \left.+\frac{2\Psi_3(\vec{k}_1,\vec{k}_2,\vec{k}_{1,3})2\Psi_4(\vec{k}_3,\vec{k}_4,\vec{k}_5,\vec{k}_{1,4})}{2k_{1,3}}\bigg \vert_{\begin{gathered}
    \begin{tikzpicture}[line width=0.6,scale=0.27,line cap=round,every node/.style={font=\scriptsize}]
        \coordinate (L1) at (-1,0);
        \coordinate (M) at (1,0);
        \coordinate (R1) at (3,0);
        \draw (L1) -- (R1);
        \filldraw[color=black,fill=black] (L1) circle[radius=0.2];
        \filldraw[color=black,fill=black] (R1) circle[radius=0.2];
        \filldraw[color=black,fill=black] (M) circle[radius=0.2];
        \draw[color=Maroon,thick] (0,-0.2) -- (0,0.2);
\end{tikzpicture}
\end{gathered}}\right. \\
       &\left.+\frac{2\Psi_3(\vec{k}_1,\vec{k}_2,\vec{k}_{1,3})2\Psi_3(\vec{k}_{1,3},\vec{k}_3,\vec{k}_{1,4})2\Psi_3(\vec{k}_{1,4},\vec{k}_4,\vec{k}_5)}{2k_{1,3}\,2k_{1,4}} \bigg \vert_{\begin{gathered}
    \begin{tikzpicture}[line width=0.6,scale=0.27,line cap=round,every node/.style={font=\scriptsize}]
        \coordinate (L1) at (-1,0);
        \coordinate (M) at (1,0);
        \coordinate (R1) at (3,0);
        \draw (L1) -- (R1);
        \filldraw[color=black,fill=black] (L1) circle[radius=0.2];
        \filldraw[color=black,fill=black] (R1) circle[radius=0.2];
        \filldraw[color=black,fill=black] (M) circle[radius=0.2];
        \draw[color=Maroon,thick] (2,-0.2) -- (2,0.2);
        \draw[color=Maroon,thick] (0,-0.2) -- (0,0.2);
\end{tikzpicture}
\end{gathered}}\right)\, ,
\end{split}
\label{eq:5ptCorr}
\end{equation}
where now we can represent each term separately diagrammatically as follows:
\begin{equation}
\Psi_5(\vec{k}_1,\vec{k}_2,\vec{k}_3,\vec{k}_4,\vec{k}_5)\bigg \vert_{\begin{gathered}
    \begin{tikzpicture}[line width=0.6,scale=0.27,line cap=round,every node/.style={font=\scriptsize}]
        \coordinate (L1) at (-1,0);
        \coordinate (M) at (1,0);
        \coordinate (R1) at (3,0);
        \draw (L1) -- (R1);
        \filldraw[color=black,fill=black] (L1) circle[radius=0.2];
        \filldraw[color=black,fill=black] (R1) circle[radius=0.2];
        \filldraw[color=black,fill=black] (M) circle[radius=0.2];
\end{tikzpicture}
\end{gathered}} = \left( \begin{gathered}
    \begin{tikzpicture}[line width=0.6,scale=0.27,line cap=round,every node/.style={font=\scriptsize},baseline={([yshift=-0.5ex]current bounding box.center)}]
        \coordinate (L1) at (-1,0);
        \coordinate (M) at (1,0);
        \coordinate (R1) at (3,0);
        \draw (L1) -- (R1);
        \filldraw[color=black,fill=black] (L1) circle[radius=0.2];
        \filldraw[color=black,fill=black] (R1) circle[radius=0.2];
        \filldraw[color=black,fill=black] (M) circle[radius=0.2];
        \draw[color=Green] (L1) circle[radius=0.5];
        \draw[color=Green] (R1) circle[radius=0.5];
        \draw[color=Green] (M) circle[radius=0.5];
       \draw[thick, Blue]
  ($(L1)!.5!(M)$) ellipse [x radius=2, y radius=0.8];
   \draw[thick, Purple]
  ($(L1)!.5!(R1)$) ellipse [x radius=3.6, y radius=1.2];
\end{tikzpicture}
\end{gathered} + \begin{gathered}
    \begin{tikzpicture}[line width=0.6,scale=0.27,line cap=round,every node/.style={font=\scriptsize},baseline={([yshift=-0.5ex]current bounding box.center)}]
        \coordinate (L1) at (-1,0);
        \coordinate (M) at (1,0);
        \coordinate (R1) at (3,0);
        \draw (L1) -- (R1);
        \filldraw[color=black,fill=black] (L1) circle[radius=0.2];
        \filldraw[color=black,fill=black] (R1) circle[radius=0.2];
        \filldraw[color=black,fill=black] (M) circle[radius=0.2];
        \draw[color=Green] (L1) circle[radius=0.5];
        \draw[color=Green] (R1) circle[radius=0.5];
        \draw[color=Green] (M) circle[radius=0.5];
       \draw[thick, Blue]
  ($(M)!.5!(R1)$) ellipse [x radius=2, y radius=0.8];
   \draw[thick, Purple]
  ($(L1)!.5!(R1)$) ellipse [x radius=3.6, y radius=1.2];
\end{tikzpicture}
\end{gathered} \right) = \frac{1}{\mathcal{P}_{1,2,3} \mathcal{P}_{1,3,4} \mathcal{P}_{1,4,5}E_t} \times\left( \frac{1}{\mathcal{P}_{1,2,3,4}} + \frac{1}{\mathcal{P}_{1,3,4,5}} \right) 
\label{eq:5pt1}
\end{equation}
where $P_{i,j,k,m}$ is the perimeter of the subpolygon anchored at vertices $i,j,k,m$ of the momentum polygon, $P_5$, so that for example $\mathcal{P}_{1,2,3,4} = k_1+k_2+k_3+k_{1,4}$. As explained in \cite{Cosmohedron}, these precisely correspond to the sum of the energies entering each tube in the tubings -- the three triangle factors upfront and the total energy correspond to the three green tubes and the purple tube, respectively, and the terms in parenthesis correspond to the remaining tubes. 

As for the terms with a single pole in \eqref{eq:5ptCorr} we have 
\begin{equation}
\begin{aligned}
&\frac{\Psi_4(\vec{k}_1,\vec{k}_2,\vec{k}_3,\vec{k}_{1,4})\Psi_3(\vec{k}_4,\vec{k}_5,\vec{k}_{1,4})}{k_{1,4}}\bigg \vert_{\begin{gathered}
    \begin{tikzpicture}[line width=0.6,scale=0.27,line cap=round,every node/.style={font=\scriptsize}]
        \coordinate (L1) at (-1,0);
        \coordinate (M) at (1,0);
        \coordinate (R1) at (3,0);
        \draw (L1) -- (R1);
        \filldraw[color=black,fill=black] (L1) circle[radius=0.2];
        \filldraw[color=black,fill=black] (R1) circle[radius=0.2];
        \filldraw[color=black,fill=black] (M) circle[radius=0.2];
        \draw[color=Maroon,thick] (2,-0.2) -- (2,0.2);
\end{tikzpicture}
\end{gathered}} = \left( \begin{gathered}
    \begin{tikzpicture}[line width=0.6,scale=0.3,line cap=round,every node/.style={font=\scriptsize},baseline={([yshift=-0.5ex]current bounding box.center)}]
        \coordinate (L1) at (-1,0);
        \coordinate (M) at (1,0);
        \coordinate (R1) at (3,0);
        \draw (L1) -- (R1);
        \filldraw[color=black,fill=black] (L1) circle[radius=0.2];
        \filldraw[color=black,fill=black] (R1) circle[radius=0.2];
        \filldraw[color=black,fill=black] (M) circle[radius=0.2];
        \draw[color=Green] (L1) circle[radius=0.5];
        \draw[color=Green] (R1) circle[radius=0.5];
        \draw[color=Green] (M) circle[radius=0.5];
       \draw[thick, Blue]
  ($(L1)!.5!(M)$) ellipse [x radius=1.8, y radius=0.8];
  \draw[color=Maroon,thick] (2.1,-0.2) -- (2.1,0.2);
\end{tikzpicture}
\end{gathered}\right) =  \frac{1}{\mathcal{P}_{1,2,3,4}\mathcal{P}_{1,2,3}\mathcal{P}_{1,3,4}} \times \frac{1}{k_{1,4}} \times \frac{1}{\mathcal{P}_{1,4,5}}, \\
&\frac{\Psi_3(\vec{k}_1,\vec{k}_2,\vec{k}_{1,3})\Psi_4(\vec{k}_{1,3},\vec{k}_3,\vec{k}_4,\vec{k}_5)}{k_{1,3}}\bigg \vert_{\begin{gathered}
    \begin{tikzpicture}[line width=0.6,scale=0.27,line cap=round,every node/.style={font=\scriptsize}]
        \coordinate (L1) at (-1,0);
        \coordinate (M) at (1,0);
        \coordinate (R1) at (3,0);
        \draw (L1) -- (R1);
        \filldraw[color=black,fill=black] (L1) circle[radius=0.2];
        \filldraw[color=black,fill=black] (R1) circle[radius=0.2];
        \filldraw[color=black,fill=black] (M) circle[radius=0.2];
        \draw[color=Maroon,thick] (0,-0.2) -- (0,0.2);
\end{tikzpicture}
\end{gathered}} = \left( \begin{gathered}
    \begin{tikzpicture}[line width=0.6,scale=0.3,line cap=round,every node/.style={font=\scriptsize},baseline={([yshift=-0.5ex]current bounding box.center)}]
        \coordinate (L1) at (-1,0);
        \coordinate (M) at (1,0);
        \coordinate (R1) at (3,0);
        \draw (L1) -- (R1);
        \filldraw[color=black,fill=black] (L1) circle[radius=0.2];
        \filldraw[color=black,fill=black] (R1) circle[radius=0.2];
        \filldraw[color=black,fill=black] (M) circle[radius=0.2];
        \draw[color=Green] (L1) circle[radius=0.5];
        \draw[color=Green] (R1) circle[radius=0.5];
        \draw[color=Green] (M) circle[radius=0.5];
       \draw[thick, Blue]
  ($(M)!.5!(R1)$) ellipse [x radius=1.8, y radius=0.8];
  \draw[color=Maroon,thick] (-0.1,-0.2) -- (-0.1,0.2);
\end{tikzpicture}
\end{gathered}\right) =  \frac{1}{\mathcal{P}_{1,2,3}} \times \frac{1}{k_{1,3}} \times \frac{1}{\mathcal{P}_{1,3,4,5}\mathcal{P}_{1,3,4}\mathcal{P}_{1,4,5}} ,
\end{aligned}
\label{eq:5pt2}
\end{equation}
and finally, the piece with two poles is simply
\begin{equation}
\frac{\Psi_3(\vec{k}_1,\vec{k}_2,\vec{k}_{1,3})\Psi_3(\vec{k}_{1,3},\vec{k}_3,\vec{k}_{1,4})\Psi_3(\vec{k}_{1,4},\vec{k}_4,\vec{k}_5)}{k_{1,3}\,k_{1,4}} \bigg \vert_{\begin{gathered}
    \begin{tikzpicture}[line width=0.6,scale=0.27,line cap=round,every node/.style={font=\scriptsize}]
        \coordinate (L1) at (-1,0);
        \coordinate (M) at (1,0);
        \coordinate (R1) at (3,0);
        \draw (L1) -- (R1);
        \filldraw[color=black,fill=black] (L1) circle[radius=0.2];
        \filldraw[color=black,fill=black] (R1) circle[radius=0.2];
        \filldraw[color=black,fill=black] (M) circle[radius=0.2];
        \draw[color=Maroon,thick] (2,-0.2) -- (2,0.2);
        \draw[color=Maroon,thick] (0,-0.2) -- (0,0.2);
\end{tikzpicture}
\end{gathered}} = \left( \begin{gathered}
    \begin{tikzpicture}[line width=0.6,scale=0.3,line cap=round,every node/.style={font=\scriptsize},baseline={([yshift=-0.5ex]current bounding box.center)}]
        \coordinate (L1) at (-1,0);
        \coordinate (M) at (1,0);
        \coordinate (R1) at (3,0);
        \draw (L1) -- (R1);
        \filldraw[color=black,fill=black] (L1) circle[radius=0.2];
        \filldraw[color=black,fill=black] (R1) circle[radius=0.2];
        \filldraw[color=black,fill=black] (M) circle[radius=0.2];
        \draw[color=Green] (L1) circle[radius=0.5];
        \draw[color=Green] (R1) circle[radius=0.5];
        \draw[color=Green] (M) circle[radius=0.5];
  \draw[color=Maroon,thick] (0,-0.2) -- (0,0.2);
  \draw[color=Maroon,thick] (2,-0.2) -- (2,0.2);
\end{tikzpicture}
\end{gathered}\right) = \frac{1}{\mathcal{P}_{1,2,3}} \times \frac{1}{k_{1,3}} \times \frac{1}{\mathcal{P}_{1,3,4}}\times\frac{1}{k_{1,4}} \times \frac{1}{\mathcal{P}_{1,4,5}}. 
\label{eq:5pt3}
\end{equation}

Once again to obtain the full $5$-point correlator one must further sum over all cyclic permutations of the answer above, to account for all the $5$ different diagrams. From the examples above it is then clear that all the terms that contribute to the correlator coming from a particular graph $\mathcal{G}$, share a set of singularities in common: the perimeter of the triangles entering the dual respective triangulation. Therefore, it is natural to factor out this part and look at all the remaining singularities -- which are precisely those captured by the graph correlahedra described in the main text. 

As for the full correlator, given the definition in \eqref{eq:BornRule}-\eqref{eq:wavefunc_mod_sq} together with the examples above, it is clear that after summing over all graphs the final expression for the correlator takes the form given in \eqref{eq:Corr}, where in particular at tree-level we have that the factors of two entering the action in \eqref{eq:wavefunc_mod_sq}, end up simply leading to an overall factor of $2$. The situation changes at one-loop as we now review. 

\subsection{One-loop correlator}
\label{app:one-loopCorr}

For the one-loop (or higher) correlators the derivation is exactly the same. Nevertheless, it is instructive to show a simple example because at one loop, unlike tree level, the factors of $2$ will not cancel and become an overall factor. To illustrate this point, consider the one-loop two-point correlation contribution coming from the bubble diagram:
\begin{align}
    \langle \Phi_1 \Phi_2\rangle \bigg  \vert_{\begin{gathered}
    \begin{tikzpicture}[line width=0.6,scale=0.27,line cap=round,every node/.style={font=\scriptsize}]
        \coordinate (L1) at (-1,0);
        \coordinate (R1) at (1,0);
        \draw (0,0) circle (1cm);
        \filldraw[color=black,fill=black] (L1) circle[radius=0.2];
        \filldraw[color=black,fill=black] (R1) circle[radius=0.2];
\end{tikzpicture}
\end{gathered}}
    = \frac{1}{2\,k_1\,2\,k_2}\Bigg(& 
    2\,\Psi_2^{(1)}(k_1,k_2) \bigg  \vert_{\begin{gathered}
    \begin{tikzpicture}[line width=0.6,scale=0.27,line cap=round,every node/.style={font=\scriptsize}]
        \coordinate (L1) at (-1,0);
        \coordinate (R1) at (1,0);
        \draw (0,0) circle (1cm);
        \filldraw[color=black,fill=black] (L1) circle[radius=0.2];
        \filldraw[color=black,fill=black] (R1) circle[radius=0.2];
\end{tikzpicture}
\end{gathered}}
    +\frac{2\,\Psi_4^{(0)}(k_1,y_2,y_2,k_2)}{2\,y_2}\bigg  \vert_{\begin{gathered}
    \begin{tikzpicture}[line width=0.6,scale=0.27,line cap=round,every node/.style={font=\scriptsize}]
        \coordinate (L1) at (-1,0);
        \coordinate (R1) at (1,0);
        \draw (0,0) circle (1cm);
        \filldraw[color=black,fill=black] (L1) circle[radius=0.2];
        \filldraw[color=black,fill=black] (R1) circle[radius=0.2];
        \draw[color=Maroon,thick] (0,0.8) -- (0,1.2);
\end{tikzpicture}
\end{gathered}}
    \nonumber \\ &+\frac{2\Psi_4^{(0)}(k_1,y_1,y_1,k_2)}{2\,y_1}\bigg  \vert_{\begin{gathered}
    \begin{tikzpicture}[line width=0.6,scale=0.27,line cap=round,every node/.style={font=\scriptsize}]
        \coordinate (L1) at (-1,0);
        \coordinate (R1) at (1,0);
        \draw (0,0) circle (1cm);
        \filldraw[color=black,fill=black] (L1) circle[radius=0.2];
        \filldraw[color=black,fill=black] (R1) circle[radius=0.2];
        \draw[color=Maroon,thick] (0,-0.8) -- (0,-1.2);
\end{tikzpicture}
\end{gathered}}
    +\frac{2\,\Psi_3^{(0)}(k_1,y_1,y_2)\,2\,\Psi_3^{(0)}(k_2,y_1,y_2)}{2\,y_1\,2\,y_2}\bigg  \vert_{\begin{gathered}
    \begin{tikzpicture}[line width=0.6,scale=0.27,line cap=round,every node/.style={font=\scriptsize}]
        \coordinate (L1) at (-1,0);
        \coordinate (R1) at (1,0);
        \draw (0,0) circle (1cm);
        \filldraw[color=black,fill=black] (L1) circle[radius=0.2];
        \filldraw[color=black,fill=black] (R1) circle[radius=0.2];
        \draw[color=Maroon,thick] (0,0.8) -- (0,1.2);
        \draw[color=Maroon,thick] (0,-0.8) -- (0,-1.2);
\end{tikzpicture}
\end{gathered}}
    \Bigg) \, ,
    \label{eq:bubble_corr}
\end{align}
where the superscript in the wavefunction coefficients indicates the loop order, with $\Psi^{(0)}$ being a tree-level coefficient. The variables $y_1$ and $y_2$ correspond to the momentum running in the respective loop edge. It is now clear from \eqref{eq:bubble_corr} that the pure wavefunction term has an overall factor of 2 which the remaining terms do not. The reason is simply because, at loop-level, when we delete a loop edge the answer turns into a single rather, as opposed to what happens when we delete a tree edge, which splits the answer into two disjoint graphs. This means that in any term where we delete an edge from a tree part of the diagram, we get the product of the wavefunction coefficients of the subgraphs to the ``left" and ``right" of the edge, each of them with the respective $2$ factor. However, for a loop edge, this is no longer the case, as we get a single lower wavefunction. Of course, after deleting a loop-edge to get a tree, any further edges deleted will not increase the factors of $1/2$. Hence why the last term in \eqref{eq:bubble_corr} has the same factor as the two preceding terms. Finally, from the above discussion, it should be clear that in the other contributions to the two-point correlator, such as the tadpole, not only the wavefunction will have the overall factor of $2$ but also the term where we delete the tadpole propagator edge.

\section{Five-Point Example}
\label{sec:AppExamples}

Here we go in detail into the five-point correlatron, depicted to the right of figure \ref{fig:5pt}. We start by giving some facet inequalities:
\begin{equation}
    X_{1,3} \geq 0\, ,\quad X_{1,3} + X_B \geq \epsilon_{\{(1,3)\}}\, ,\quad X_{1,4} + X_B \geq \epsilon_{\{(1,4)\}}\, ,\quad X_{1,3} + X_{1,4} + X_B \geq \epsilon_{\{(1,3),(1,4)\}}\, ,\quad X_B \geq \epsilon_{\{T\}}\, ,
    \label{eq:facet_ex}
\end{equation}
where the inequalities correspond to the facets $(X_{1,3})$, $(P_{123},P_{1345})$, $(P_{1234},P_{145})$, $(P_{123},P_{134},P_{145})$ and $(P_{12345})$, respectively, where, for example, $P_{1234}$ denotes the quadrilateral anchored at vertices $1$, $2$, $3$ and $4$. To the top facet, we associate the inequality $X_T\geq0$, which after imposing the condition $X_T+X_B=H$, is simply $H-X_B\geq0$. To obtain the correct embedding the $\epsilon_C$ must obey inequalities as we saw in the main text (at five points there are no equalities). For example, the $\epsilon_C$ in \eqref{eq:facet_ex} must obey the inequalities:
\begin{equation}
    \epsilon_{\{(1,3)\}} + \epsilon_{\{(1,4)\}} < \epsilon_{\{(1,3),(1,4)\}}\, , \quad  \epsilon_{\{(1,3)\}} < \epsilon_{\{(1,3),(1,4)\}}\, ,\quad \epsilon_{\{(1,4)\}} < \epsilon_{\{(1,3),(1,4)\}}\, ,
\end{equation}
where the first inequality comes from \eqref{eq:cosmo_eps_ineqs}, and the other two are examples of \eqref{eq:new_eps_ineqs}. A simple way to parameterize the $\epsilon_C$ such that all inequalities are satisfied is by using the map \eqref{eq:eps_to_delta_map}, which turns the above inequalities into:
\begin{equation}
    \delta_{1234} + \delta_{2345} < \delta_{134} + \delta_{12345}\, , \quad \delta_{2345} < \delta_{134} + \delta_{145}\, , \quad \delta_{1234} < \delta_{123} + \delta_{134}\, ,
\end{equation}
where the first inequality is of the form of the first inequality in \eqref{eq:delta_ineqs}, and the other two inequalities are of the form of the second inequality in \eqref{eq:delta_ineqs}, which are the new type of inequalities with respect to the cosmohedron inequalities. It is then clear that the parameterization of the $\delta$'s given in \eqref{eq:delta_param} satisfies the inequalities above. Following these same steps with the remaining facets yields the correct embedding of the full correlatron. 

Let us now explain how we can extract the correlator from it. To do this we will give some examples of what happens in the vertices highlighted in yellow (vertex 3), green (vertex 1), and purple (vertex 2) in figure \ref{fig:5pt}. Note that since all these vertices lie on the facet $(P_{123},P_{134},P_{145})$, the terms we obtain will be associated with contributions to the correlator coming from the graph with propagators $k_{1,3}$ and $k_{1,4}$, which we described explicitly in \eqref{eq:5pt1}-\eqref{eq:5pt2}-\eqref{eq:5pt3}. Following the prescription given in the main text, we start by  attributing to each of the facets in \eqref{eq:facet_ex} the following factors:
\begin{align}
    &(X_{1,3}): \frac{1}{k_{13}}\, , \quad (P_{123},P_{1345}): \frac{1}{\mathcal{P}_{123}\,\mathcal{P}_{1345}}\, , \quad (P_{1234},P_{145}): \frac{1}{\mathcal{P}_{145}\,\mathcal{P}_{1234}}\, , \nonumber  \\ 
    &(P_{123},P_{134},P_{145}): \frac{1}{\mathcal{P}_{123}\,\mathcal{P}_{134}}+\frac{1}{\mathcal{P}_{134}\,\mathcal{P}_{145}}+\frac{1}{\mathcal{P}_{123}\mathcal{P}_{134}\,\mathcal{P}_{145}}\, , \quad (P_{12345}): \frac{1}{E_T}\, ,
    \label{eq:sings_fivePts}
\end{align}
where in this case all facets have at most one non-triangle subpolygon and therefore $n_C=1$, in all facets. Note that while we are using $P_{\{i\}}$ to denote the subpolygon, $\mathcal{P}_{\{i\}}$ stands for the respective perimeter variable associated with it. 

Let's now compute the contribution from the vertex which is the intersection of the facets $(P_{12345})$, $(P_{123},P_{134},P_{145})$ and $(P_{123},P_{1345})$, the yellow vertex in figure \ref{fig:5pt}, vertex 3. Then we take the product of all three corresponding factors from \eqref{eq:sings_fivePts}, to obtain:
\begin{equation}
    \frac{1}{E_T\,\mathcal{P}_{123}^2\,\mathcal{P}_{134}\,P_{1345}}+\frac{1}{E_T\,\mathcal{P}_{123}\,\mathcal{P}_{134}\,\mathcal{P}_{145}\,P_{1345}}+\frac{1}{E_T\,\mathcal{P}_{123}^2\mathcal{P}_{134}\,\mathcal{P}_{145}\,P_{1345}} \to \frac{1}{E_T\,\mathcal{P}_{123}\,\mathcal{P}_{134}\,\mathcal{P}_{145}\,P_{1345}}\, ,
    \label{eq:yellowvertex}
\end{equation}
where in $\to$ we drop all terms with higher poles. This precisely gives us a contribution to the correlator from a term that comes from the pure wavefunction part (as expected since this vertex lies in the cosmohedron facet), as one can check by looking at \eqref{eq:5pt1}. Let us now look at the vertex coming from the intersection of the facets $(X_{1,3})$, $(P_{123},P_{134},P_{145})$ and $(P_{123},P_{1345})$ (the green vertex in figure \ref{fig:5pt}, vertex 1), in this case taking the product of the factors associated to each facet yields:
\begin{equation}
    \frac{1}{k_{1,3}\,\mathcal{P}_{123}^2\,\mathcal{P}_{134}\,P_{1345}}+\frac{1}{k_{1,3}\,\mathcal{P}_{123}\,\mathcal{P}_{134}\,\mathcal{P}_{145}\,P_{1345}}+\frac{1}{k_{1,3}\,\mathcal{P}_{123}^2\mathcal{P}_{134}\,\mathcal{P}_{145}\,P_{1345}} \to  \frac{1}{k_{1,3}\,\mathcal{P}_{123}\,\mathcal{P}_{134}\,\mathcal{P}_{145}\,P_{1345}}\, ,
    \label{eq:greenvertex}
\end{equation}
which is identical to the contribution in \eqref{eq:yellowvertex} (from the green vertex), but now we have a singularity in $k_{1,3}$, instead of the total energy, $E_T$, which precisely matches the term in the bottom equation in \eqref{eq:5pt2}. Finally, to emphasize the difference between the prescription for wavefunction and the correlator, let us look at the vertex which is the intersection of the facets $(X_{1,3})$, $(P_{123},P_{134},P_{145})$ and $(X_{1,4})$ (the violet vertex in figure \ref{fig:5pt}, vertex 2), where the new facet, $(X_{1,4})$, comes with a factor of $1/(k_{1,4})$. The contribution from this vertex is:
\begin{equation}
    \frac{1}{k_{1,3}\,k_{1,4}\,\mathcal{P}_{123}\,\mathcal{P}_{134}}+\frac{1}{k_{1,3}\,k_{1,4}\,\mathcal{P}_{134}\,\mathcal{P}_{145}}+\frac{1}{k_{1,3}\,k_{1,4}\,\mathcal{P}_{123}\mathcal{P}_{134}\,\mathcal{P}_{145}} \to \frac{1}{k_{1,3}\,k_{1,4}\,\mathcal{P}_{123}\mathcal{P}_{134}\,\mathcal{P}_{145}}\, ,
\end{equation}
where we see that there is no term with higher degree poles, so instead in $\to$ we throw away all terms with the wrong energy units, which in this case is $|\vec{k}|^{-5}$. Doing this precisely leaves us with the desired term given in \eqref{eq:5pt3}.

To extract the wavefunction at five-points we did not need to simplify the polytope since it is simple to begin with. At higher points, such as six points, there will be vertices that are non-simple. In order to extract the answer we need first to simplify these vertices by turning each equality into an inequality of the form \eqref{eq:cosmo_eps_ineqs} (turning the correlatron into the permuto-correlatron), and attribute the factors to each facet according to \eqref{eq:WF_permuto_sings}. After doing this, we proceed exactly in the way described above to compute the contribution from all the vertices of the permuto-correlatron, which finally gives us the full correlator.

\section{Graph Correlahedron as  a Canonical Form}
\label{sec:PhysKinGraphCorr}
\begin{figure}[t]
    \centering
    \includegraphics[width=\linewidth]{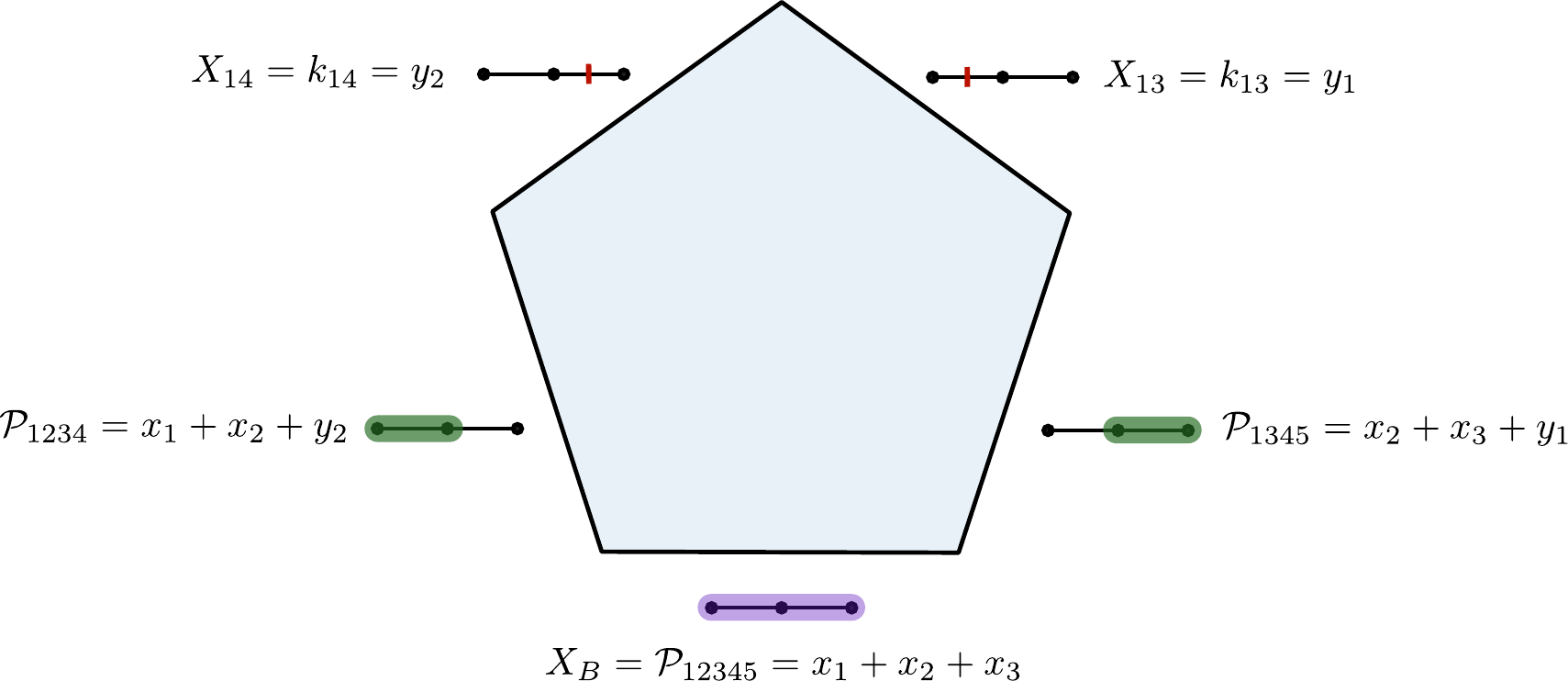}
    \caption{$5$-point graph correlahedron for the graph containing propagators $\{(1,3),(1,4)\}$ with respective facet labels and the map to the usual $x$ and $y$ energy variables.}
    \label{fig:5ptGraphCorr}
\end{figure}
In this appendix we discuss in detail the $5$-point graph correlahedron to show how factoring out the common triangle singularities lets us write it as a canonical form, directly in physical kinematics -- provided we use the correct embedding. 

Without loss of generality, let us choose the underlying triangulation of the pentagon to be $\mathcal{T} =\{(1,3),(1,4)\}$. As explained in the main text and seen explicitly in appendix \ref{sec:AppCorrelator}, all terms in the graph correlator share a set of singularities corresponding to the perimeter of the triangles in $\mathcal{T}$:
\begin{equation}
    \mathcal{P}_{123} = k_1 +k_2 +k_{1,3} = x_1 + y_1, \quad \mathcal{P}_{134} = k_{1,3} +k_4 +k_{1,4} = y_1 +x_2 + y_2, \quad \mathcal{P}_{145} = k_{1,4} +k_4 +k_5 =  y_2 + x_3,
    \label{eq:frozenTs}
\end{equation}
where here we introduce the map between the perimeters, the $k$'s and the $x/y$ variables which parametrize physical kinematics as follows $y_1=k_{1,3}$, $y_2=k_{1,4}$ and $x_1=k_1+k_2$, $x_2=k_3$, $x_3=k_4+k_5$.

So if we factor out these singularities we are left with the following expression:
\begin{equation}
\begin{aligned}
    \langle \Phi_1 \cdots \Phi_5 \rangle_{\mathcal{T}}  &\propto \frac{1}{k_{1,3} k_{1,4}} + \frac{1}{k_{1,3} \mathcal{P}_{1345}}+ \frac{1}{\mathcal{P}_{1345} \mathcal{P}_{12345}}+\frac{1}{\mathcal{P}_{1234} \mathcal{P}_{12345}} +\frac{1}{\mathcal{P}_{1234} k_{1,4}} \\
    & \propto \frac{1}{y_1 y_2} + \frac{1}{y_1 (x_2+x_3+y_1)}+ \frac{1}{(x_2+x_3+y_1)(x_1+x_2+x_3)} + \frac{1}{(x_1+x_2+y_2)(x_1+x_2+x_3)} \\
    & \quad + \frac{1}{(x_1+x_2+y_2)y_2},
\end{aligned}
    \label{eq:GraphCorr5}
\end{equation}
which has $5$ different poles, each paired in two to give a term of the graph correlator. It is then natural to associate each of these pole factors to an edge of a pentagon, as depicted in figure \ref{fig:5ptGraphCorr}, so that each vertex represents a term of the correlator. This combinatorially gives us the $5$-point graph correlahedron. 

To embed this pentagon in physical kinematics, we can start by defining the positive region in $(x_1,x_2,x_3,y_1,y_2)$ space corresponding to 
\begin{equation}
\begin{aligned}
     &\, y_1\geq 0, \quad  y_2\geq 0, \quad (x_1+x_2+x_3)  \geq 0,\\
     &(x_1+x_2+y_2) \geq 0,\quad (y_1+x_2+x_3)  \geq 0,
\end{aligned} \quad \quad \text{or} \quad \quad \begin{aligned}
     &\, k_{1,3}\geq 0, \quad  k_{1,4}\geq 0, \quad \mathcal{P}_{12345}  \geq 0,\\
     &\mathcal{P}_{1234} \geq 0,\quad \mathcal{P}_{1345}  \geq 0,
\end{aligned}
\label{eq:PhysEmbedd}
\end{equation}
where we also give the equivalent set of inequalities in $(k_{1,3},k_{1,4},\mathcal{P}_{1234},\mathcal{P}_{1345},\mathcal{P}_{12345})$ space. In both cases, we have that all boundaries of the space are associated to a singularity of \eqref{eq:GraphCorr5}. However, these inequalities define a $5$-dimensional space, while the pentagon (in figure \ref{fig:5ptGraphCorr}) is $2$-dimensional. So to correctly reduce the dimension of this positive region, we intersect it with some plane. One way of doing this is by considering the plane where the triangle perimeters \eqref{eq:frozenTs} are fixed and positive, this is a natural choice since we are factoring out these singularities so it is as if they are effectively frozen. The plane of constant triangles corresponds to 
\begin{equation}
\mathbb{T} = \begin{cases}
    2k_{1,3} + 2k_{1,4} + \mathcal{P}_{12345} = t_1 + t_2 +t_3, \\   
    2k_{1,3} + \mathcal{P}_{1234} = t_1 + t_2, \\
    2k_{1,4} + \mathcal{P}_{1345} = t_2 + t_3, \\
\end{cases}
\label{eq:trianglePlane}
\end{equation}
where $\mathcal{P}_{123}=t_1$ , $\mathcal{P}_{134} =t_2$ and $ \mathcal{P}_{145}=t_3$, are positive constants. Using this, we can write $(\mathcal{P}_{12345},\mathcal{P}_{1234},\mathcal{P}_{1345})$ in terms of $k_{1,3}$ and $k_{1,4}$; or equivalently write $(x_1,x_2,x_3)$ in terms of $y_1$ and $y_2$. By replacing this into \eqref{eq:GraphCorr5}, we get in $(y_1,y_2) \equiv (k_{1,3},k_{1,4})$ space
\begin{equation}
\begin{aligned}
 \langle \Phi_1 \cdots \Phi_5 \rangle_{\mathcal{T}} \bigg \vert_{\mathbb{T}}   \propto  &\frac{1}{y_1 y_2} + \frac{1}{y_1 (t_2+t_3 -2y_2)}+ \frac{1}{(t_2+t_3 -2y_2)(t_1+t_2+t_3-2y_1-2y_2)} \\
    &+ \frac{1}{(t_1+t_2-2y_1)(t_1+t_2+t_3-2y_1-2y_2)}  + \frac{1}{(t_1+t_2-2y_1)y_2}.
\end{aligned}
\end{equation}

Now one can easily check that the expression above is \textit{not} a canonical form in $(y_1,y_2)$ space. In particular, some the maximal residues are not $\pm 1$. The existence of the non-unit residues is what leads to the introduction of weighted polytopes when describing geometries for the correlator such as in \cite{GeometryCosmoCorr}. The expression above should be what one obtains from the weighted polytope in \cite{GeometryCosmoCorr}, by slicing it through the plane where the triangle perimeters are fixed. This is the same procedure that takes us from the graph associahedra for the wavefunction to the cosmological polytopes. 

However, what we want to show, is that quite remarkably, there is a \textit{different} slicing of the original $5$-dimensional space in \eqref{eq:PhysEmbedd}, naturally given by the correlatron, which leaves us with a $2$-d space whose \textit{canonical form} is exactly the correlator! From the correlatron we have that $k_{1,3} = X_{1,3} , \quad  k_{1,4} = X_{1,4}, \quad  \mathcal{P}_{12345} = X_{B}$, and the slicing which gives the graph correlahedron is instead \footnote{here we are assuming for simplicity that $\delta_{12345}=0$}:
\begin{equation}
\mathbb{C} = \begin{cases}
 X_{1,3}+X_{1,4}+X_{B} = \delta_{123} + \delta_{134} +\delta_{145},\\
    \mathcal{P}_{1234} = X_{1,4} + X_{B} - \delta_{145} - \delta_{1234} ,\\
    \mathcal{P}_{1345} = X_{1,3} + X_{B} -\delta_{123} -\delta_{1345}, \\
    \end{cases} \quad \Leftrightarrow\quad  \mathbb{C} =\begin{cases}
    k_{1,3}+k_{1,4}+\mathcal{P}_{12345} = \delta_{123} + \delta_{134} +\delta_{145}\\
   k_{1,3} + \mathcal{P}_{1234} = \delta_{123} +\delta_{134} - \delta_{1234} ,\\
     k_{1,4}+\mathcal{P}_{1345} =\delta_{134}+ \delta_{145} - \delta_{1345}, 
    \end{cases}
\end{equation}
with the $\delta$'s satisfying \eqref{eq:delta_ineqs}. On the right, we express the correlatron slicing directly in terms of the original variables to make manifest that this slicing is \textit{not} the same as freezing the triangle variables (given in \eqref{eq:trianglePlane}). Nonetheless, it is remarkably close to it! The only difference is that we have $2k_{1,j} \to k_{1,j}$. Now using the relations above we can write again $(\mathcal{P}_{12345},\mathcal{P}_{1234},\mathcal{P}_{1345})$ in terms of $(k_{1,3},k_{1,4}) \equiv(y_1,y_2)$, and check that the correlator \eqref{eq:GraphCorr5} in this space is indeed a canonical form! 

So quite remarkably, we find that if we factor out the triangle singularities from the graph correlator, the remaining function is a canonical form directly in physical kinematic space! However the correct embedding is not given by freezing the triangle variables, but instead that given by the correlatron, which gives the same equations but where $2y_i \to y_i$. 

One can easily check that this construction holds at higher points. For a general graph at general $n$, specified by some triangulation $\mathcal{T}_n$, when we fix all the triangles $t \subset \mathcal{T}_n$, for a given subpolygon (tube) $P$, we have the following constraint on its perimeter: 
\begin{equation}
    \mathcal{P}_P =  -2 \sum_{(i,j) \subset P} k_{i,j} +\sum_{t \subset P} \mathcal{P}_t
    \label{eq:gentriangleSlicing}
\end{equation}
where the first sum is over the chords $(i,j)$ corresponding to the propagators that live inside the $P$ (note that this does not include the chords that are boundaries of $P$), and the second sum over all triangles of $\mathcal{T}_n$ that are contained inside $P$. These equations for all subpolygons $P$, as well as for the full $P_n$, give the $n$-point analog of the $\mathbb{T}$ slicing described in \eqref{eq:trianglePlane} for $n=5$ -- for which the correlator is \textit{not} a canonical form. 

On the other hand, from the correlatron, given a subpolygon, we have the facet inequality given in \eqref{eq:genFacetIneq}, which corresponds to the slicing
\begin{equation}
    \mathcal{P}_P = \sum_{(i,j) \nsubseteq P} X_{i,j} +X_B - \sum_{t \nsubseteq P} \delta_t -  \delta_P 
    \label{eq:slicing1}
\end{equation}
where the sums are over all chords/ all triangles \textit{not} contained in $P$, and recall $X_{i,j} \equiv k_{i,j}$ and $X_B \equiv \mathcal{P}_{P_n}$. However, since the graph correlahedron is defined under the equality:
\begin{equation}
    X_B + \sum_{(i,j) \in \mathcal{T}_n} X_{i,j} = \sum_{t \subset \mathcal{T}_n} \delta_t \quad \Leftrightarrow \quad  \mathcal{P}_{P_n} = -\sum_{(i,j)\subset P_n} k_{i,j} +\sum_{t \subset P_n} \delta_t , 
    \label{eq:slic1}
\end{equation}
we can rewrite \eqref{eq:slicing1} as 
\begin{equation}
     \mathcal{P}_P = -\sum_{(i,j) \subset P} k_{i,j}  + \sum_{t \subset P} \delta_t -  \delta_P,
     \label{eq:slic2}
\end{equation}
which is exactly the same slicing as in \eqref{eq:gentriangleSlicing} but with $2k_{i,j} \to k_{i,j}$. Finally for the perimeters, $\mathcal{P}_P$, to be defined under physical kinematics, $i.e.$ as perimeters of subpolygons of the underlying momentum $n$-gon, they must satisfy:
\begin{equation}
    \mathcal{P}_P + \mathcal{P}_{P^\prime} = \mathcal{P}_{P\cup P^\prime} + \mathcal{P}_{P\cap P^\prime},
    \label{eq:PhysKin}
\end{equation}
where $P$ and $P^\prime$ are two non-triangle subpolygons that have a non-trivial overlap, $i.e.\,  P\cap P^\prime \neq \emptyset$. Just like for the cosmohedron \cite{Cosmohedron}, these imply at the level of the $\delta_P$'s that they must satisfy:
\begin{equation}
\begin{cases}
    \delta_P + \delta_{P^\prime} < \delta_{P \cup P^\prime} + \delta_{P \cap P^\prime},
\quad  {\rm if \,} P \cap P^\prime \, {\rm is \, a \, triangle} , \\
\delta_P + \delta_{P^\prime} = \delta_{P \cup P^\prime} + \delta_{P \cap P^\prime},
\quad  {\rm otherwise}\, .
\end{cases}
\label{eq:ineqsDegGA}
\end{equation}

By imposing these relations between the $\delta$'s, together with the slicings \eqref{eq:slic1}-\eqref{eq:slic2},  we have defined a polytope in physical kinematics, $(k_{i,j}\in \mathcal{T}_n)$ space, whose canonical form is the graph correlator.

\section{One-loop correlatron}
\label{app:one-loop}

\par
\begin{figure*}[t]
    \centering
    \includegraphics[width=\linewidth]{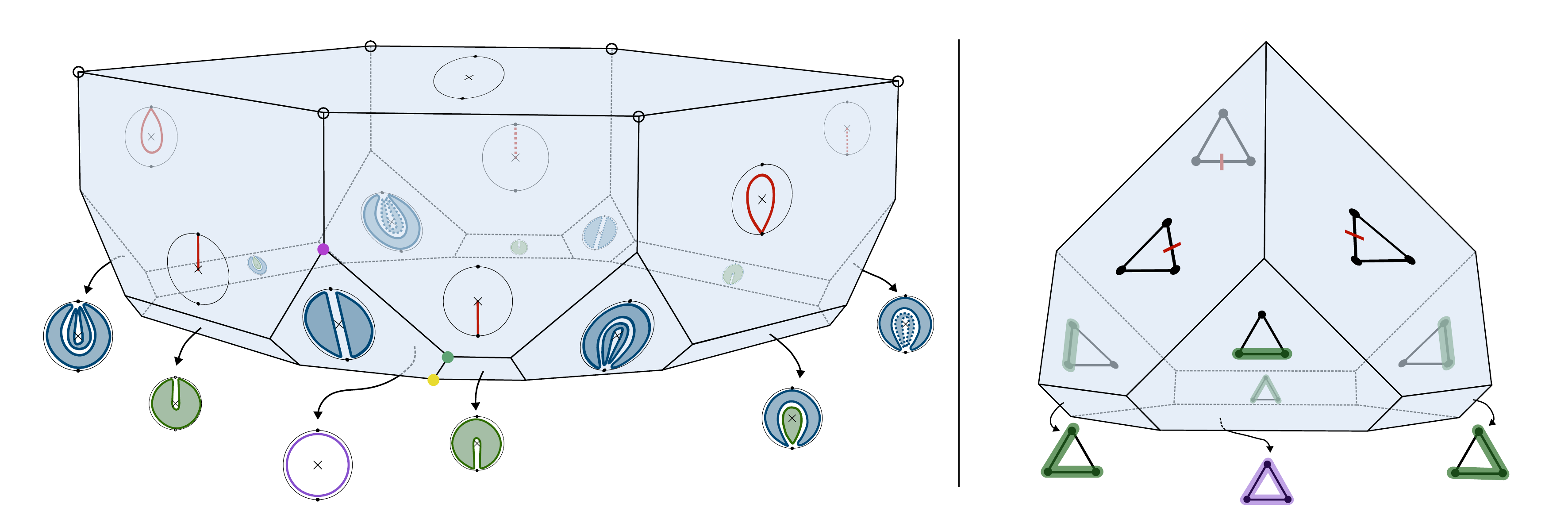}
    \caption{\textbf{Left} $2$-point one-loop Correlatron, with the respective facet labellings. \textbf{Right} Example of loop graph correlahedron for the one-loop triangle diagram.}
    \label{fig:LoopCorr}
\end{figure*}

 The generalization of the ABHY associahedron to one-loop Tr$[\phi^3]$ integrands was shown in~\cite{Arkani-Hamed:2019vag,BinGeom,Arkani-Hamed:2023mvg,Arkani-Hamed:2023lbd}, and the all-loop extension is given by surfacehedra~\cite{Surfacehedra}. For the wavefunction, the cosmohedron picture was similarly extended for all loops using associahedra and surfacehedra as the base polytopes~\cite{Cosmohedron}. It is then very natural to expect the construction of the loop correlatron to also follow straightforwardly. Indeed, the loop polytope for the correlator is also a ``sandwich'' between the amplitude polytope, and the wavefunction polytope, and its' embedding follows exactly the same structure as tree level (see left of figure \ref{fig:LoopCorr} for the example at one-loop two-points). The facet inequalities are given by \eqref{eq:facet_ineqs}, with the difference that instead of chords in a polygon, we have curves on the punctured disk. Using the generalized kinematics proposed in \cite{Salvatori:2018aha,SurfaceKin}, we consider all curves that are not homotopic to each other, and assign a different variable to each of these, such that a curve from $i$ to $j$ going through the left of the puncture, $X_{i,j}$, is assigned a different variable than the one going through the right, $X_{j,i}$. In addition, at loop-level (and in particular at one-loop), we also have the tadpole variables $X_{i,i}$, and loop variables ending on the puncture $p$, $X_{i,p}$, and $\tilde{X}_{i,p}$ \footnote{The doubling of the variables ending on the puncture, $X_{i,p}$ and $\tilde{X}_{i,p}$, is important to obtain a closed polytope at one-loop as explained in \cite{Surfacehedra}.}.

So just like at tree-level the rays of the correlatron fan/its facets are labeled by either single chords (corresponding to the associahedron facets) or collections of curves,  $\mathcal{C}$, which correspond to subsurface tilings of the punctured disk (cosmohedra facets). To the latter, we associate the inequalities in \eqref{eq:cosmo_facets}, and the $\epsilon_{\mathcal{C}}$ must obey the same equalities and inequalities as in \eqref{eq:cosmo_eps_ineqs} and \eqref{eq:new_eps_ineqs}. The map \eqref{eq:eps_to_delta_map} once more guarantees that all equalities are automatically satisfied, and similarly for inequalities as long as $\delta_{P_n}=0$ (where $P_n$ now stands for the punctured disk with $n$-points in the boundary) and the remaining $\delta_P$, for $P$ some subsurface of $P_n$, satisfy:
\begin{equation}
    \delta_P + \delta_{P^\prime} < \delta_{P \cup P^\prime} + \sum_{\tilde{P}\in\{P \cap P^\prime \}}\delta_{\tilde{P}}\, ,
    \label{eq:deltaIneqsLoop}
\end{equation}
where the sum on the RHS reflects the fact that, at loop level, the intersection of two surfaces can be a set of disjoint surfaces.  

As for tree-level, we have two extra facets:  the loop-cosmohedron as the bottom facet (labeled by the subsurface tilling given by the full disk, represented in purple on the left of figure \ref{fig:LoopCorr}), and the loop-associahedron in the top facet (which of course only exists in the truncated version of the polytope).

In figure \ref{fig:LoopCorr}, we label the facets associated with subsurface tilings  (from the loop-cosmohedron) with blue and green contours, where the cases bounded by solid lines correspond to collections involving $X_{i,p}$, while those bounded by dashed lines to collections involving $\tilde{X}_{i,p}$ variables. The single-chord facets (from the loop-associahedron), are labeled by the red curves, and once again the solid lines correspond to $X_{i,p}$ variables and the dashed ones to  $\tilde{X}_{i,p}$ variables. 

The facet factorization structure of the one-loop correlatron follows the same rule as tree-level, as we can check in figure \ref{fig:LoopCorr}. For example, the front facet in \ref{fig:LoopCorr} is associated with the curve connecting the bottom boundary point to the puncture, which cuts the punctured disk into a disk with $4$-points in the boundary -- the $4$-point tree-level surface -- and thus this facet is precisely given by the tree four-point correlatron (a hexagon). As for the subsurface tiling facets, we again find that they are given by the product of the graph correlahedra obtained by inserting a node in the middle of each surface and the cosmohedra of the respective sub-surfaces. 

In addition, the facets of the loop correlatron given by subsurface tilings where all subsurfaces are $3$-point (specifying a triangulation, $\mathcal{T}$, of the punctured disk), give us the one-loop graph-correlahedra -- the polytopes which capture the terms contributing to correlator from the graph dual to $\mathcal{T}$.  For example, on the right of figure \ref{fig:LoopCorr} we see an example of a graph correlahedra at three points, thus this is a facet of the one loop three point correlatron.

Finally, to extract the one-loop correlator from the correlatron, there are important differences with respect to tree level, some of which we have already seen in the cosmohedron \cite{Cosmohedron}. Firstly, we simplify the polytope by turning every equality into an inequality of the form \eqref{eq:cosmo_eps_ineqs}. Then, to the associahedron facets, labeled by a single cords $(i,j)$ (excluding loop variables $(i,p)$), we associate a factor:
\begin{equation}
    \frac{1}{X_{i,j}}\equiv \frac{1}{k_{i,j}}\, .
\end{equation}
For loop variables, we associate the factor:
\begin{equation}
    \frac{1}{X_{i,p}}\equiv \frac{1}{\tilde{c} \, y_{i}}\, ,
\end{equation}
where $y_{i}$ is the norm of the momenta running in the loop edge, and $\tilde{c}$ will be important to restore the correct factors of $1/2$ (described in \ref{app:one-loopCorr}), as we explain momentarily.  
For the facets associated with sub-surfaces, in other words, the facets touching the cosmohedron facet, we associate a factor of the type:
\begin{equation}
\frac{1}{X_\mathcal{C}} \equiv \frac{1}{n_C}\left[ \sum_{(P|P^\prime) \subset \mathcal{C}} \frac{1}{{\cal P}_P {\cal P^\prime}_{P^\prime}} + \prod_{P \subset \mathcal{C}}\frac{1}{\mathcal{P}_P}\right].
\label{eq:WF_loop_permuto_sings}
\end{equation}
This is not quite the same case as tree level because, just like for the loop cosmohedron, we need to associate a triangle sub-surface, $t_{(i,p)}$, to the cord $(i,p)$. For facets which have a sub-surface with a border being $(i,p)$, in \eqref{eq:WF_loop_permuto_sings}, there will be a singularity with the pair $t_{(i,p)}$ and the sub-surface which borders it, and the last term in \eqref{eq:WF_loop_permuto_sings}, which contains all the sub-surface singularities in the denominator must also have $t_{(i,p)}$. 

Finally, given these facet factors, we compute the canonical form of the permuto-correlatron keeping only terms with single poles and the correct units. In addition at loop-level we need to do two extra things, first, we set all $t_{(i,p)}\to 1$, and finally to correctly restore the factors of $2$, we want to effectively replace all $\tilde{c}^n$ by a factor of $1/2$. Recall that each $\tilde{c}$ appears in the case where we have deleted a loop-edge and so all such terms should have an overall $1/2$, with respect to the remaining. This can be done in the following way
\begin{equation}
    \langle \Phi_1 \Phi_2 \cdots \Phi_n \rangle_{\text{1-loop}} =  \mathop{\mathrm{Res}}_{\tilde{c}=0} \left[ \left(\frac{1}{\tilde{c}} + \frac{1}{2} \frac{1}{1-\tilde{c}} \right)\widetilde{\Omega} \left[\text{PermCorr}_{n} \right] (\tilde{c})  \right].
    \label{eq:Res1loop}
\end{equation}
as this residue correctly ensures that the piece with no $\tilde{c}$ comes with a factor of $1$, while the terms with any power of $\tilde{c}$ in the denominator come with a factor of $1/2$; and we have used  $\widetilde{\Omega} $ to denote
\begin{equation}
   \widetilde{\Omega} \left[\text{PermCorr}_{n} \right] \equiv  \Omega\left[\text{PermCorr}_{n} \right] \vert_{\text{single poles}, |k|^{-2n}, t_{(i,p)} \to 1}.
\end{equation}

\subsection{Example: 2-points one-loop}

Let us now consider some examples for the two-point one loop correlahedron. Just like for tree level, we start by listing some facet inequalities:
\begin{equation}
    X_{1,p}\geq 0\, , \quad X_{2,p}\geq 0\, , \quad X_{1,p} + X_B\geq \epsilon_{\{(1,p)\}}\, ,\quad X_{2,p} + X_B\geq \epsilon_{\{(2,p)\}}\, ,\quad X_{1,p}+ X_{2,p} + X_B\geq \epsilon_{\{(1,p),(2,p)\}}\, , 
    \label{eq:facet_ineqs_loop}
\end{equation}
where once more the loop cosmohedron facet is $X_B\geq 0$. In this case, the full triangulation listed above (last inequality in \eqref{eq:facet_ineqs_loop}) corresponds to the bubble diagram.
Our one-loop two point correlatron is a simple polytope, therefore there are no equalities associated to the $\epsilon$. But, in order to obtain the correct embedding we still need to impose inequalities between them. Concretely for the $\epsilon_\mathcal{C}$ in \eqref{eq:facet_ineqs_loop}, we obtain the following inequality (already present for the loop-cosmohedron):
\begin{equation}
    \epsilon_{\{(1,p)\}} + \epsilon_{\{(2,p)\}} < \epsilon_{\{(1,p),(2,p)\}} \, .
\end{equation}
Additionally, the new inequalities needed for the correlatron are:
\begin{equation}
    \epsilon_{\{(1,p)\}}  < \epsilon_{\{(1,p),(2,p)\}} \, , \quad \epsilon_{\{(2,p)\}}  < \epsilon_{\{(1,p),(2,p)\}}.
\end{equation}
Now we can again use the map from $\epsilon_C$ to $\delta_P$ in \eqref{eq:eps_to_delta_map}  to rewrite the inequalities above into:
\begin{equation}
\begin{aligned}
        \delta_{\{(1,2),(2,1),(1,p)\}}+&\delta_{\{(1,2),(2,1),(2,p)\}} < \,\delta_{\{(1,2),(1,p),(2,p)\}}+ \delta_{\{(2,1),(1,p),(2,p)\}}\, , \\
        &\delta_{\{(1,2),(2,1),(1,p)\}} <\, \delta_{\{(1,2),(1,p),(2,p)\}}+ \delta_{\{(2,1),(1,p),(2,p)\}} \,,\\
      &\delta_{\{(1,2),(2,1),(2,p)\}} < \,\delta_{\{(1,2),(1,p),(2,p)\}}+ \delta_{\{(2,1),(1,p),(2,p)\}} \, ,
\end{aligned}
\end{equation}
where we specify the subsurface associated to $\delta$ by giving the curves bounding the subsurface.
Finally, once we get the correct embedding, the last step is to extract the answer from the polytope. The singularities we attribute to each facet in \eqref{eq:facet_ineqs_loop} are the following:
\begin{equation}
\begin{aligned}
    (X_{1,p}): \frac{1}{\tilde{c}\, y_1}\, , \quad \quad (X_{2,p}):& \frac{1}{\tilde{c}\, y_2}\,  ,\quad \quad  (X_B):\frac{1}{E_T}\,, \\
    (P_{(1,2),(2,1),(1,p)}): \frac{1}{\mathcal{P}_{(1,2),(2,1),(1,p)} t_{(1,p)}}\, , \quad &\quad (P_{(1,2),(2,1),(2,p)}): \frac{1}{\mathcal{P}_{(1,2),(2,1),(2,p)} t_{(2,p)}} \\
    (P_{(1,2),(1,p),(2,p)}, \,P_{(2,1),(1,p),(2,p)}): &\frac{1}{\mathcal{P}_{(1,2),(1,p),(2,p)} \mathcal{P}_{(2,1),(1,p),(2,p)}} \,.
\end{aligned}
\end{equation}
where the perimeter variables are given by:
\begin{equation}
    \mathcal{P}_{(1,2),(2,1),(j,p)} = k_1 +k_2 +y_j, \quad \mathcal{P}_{(1,2),(1,p),(2,p)} = k_1 +y_1 +y_2, \quad \mathcal{P}_{(2,1),(1,p),(2,p)} = k_2 +y_1 +y_2.
\end{equation}
With the singularities for each facet, we can compute the contributions from the vertices marked in yellow, green, and purple in figure \ref{fig:LoopCorr}. Starting with the yellow vertex, we just take the product of every facet that touches it:
\begin{equation}
    \frac{1}{E_T \mathcal{P}_{(1,2),(1,p),(2,p)} \mathcal{P}_{(2,1),(1,p),(2,p)} \mathcal{P}_{(1,2),(2,1),(1,p)} t_{(1,p)}} \to \frac{1}{E_T \mathcal{P}_{(1,2),(1,p),(2,p)} \mathcal{P}_{(2,1),(1,p),(2,p)} \mathcal{P}_{(1,2),(2,1),(1,p)}} \, ,
\end{equation}
where in $\to$ we set $t_{(1,p)}$ to one as prescribed. Then, the contribution from the green vertex is:
\begin{equation}
    \frac{1}{\tilde{c}\, y_1 \mathcal{P}_{(1,2),(2,1),(1,p)} t_{(1,p)} \mathcal{P}_{(1,2),(1,p),(2,p)} \mathcal{P}_{(2,1),(1,p),(2,p)}} \to  \frac{1}{2 y_1 \mathcal{P}_{(1,2),(2,1),(1,p)} \mathcal{P}_{(1,2),(1,p),(2,p)} \mathcal{P}_{(2,1),(1,p),(2,p)}}\, ,
\end{equation}
where now set $t_{(1,p)}$ to one, but also after computing the residue in  $\tilde{c}$ given in \eqref{eq:Res1loop}, we obtain the correct factor of $1/2$. Finally, computing the contribution from the violet vertex yields:
\begin{equation}
    \frac{1}{\tilde{c}^2 \, y_1 \, y_2 \mathcal{P}_{(1,2),(1,p),(2,p)} \mathcal{P}_{(2,1),(1,p),(2,p)}} \to   \frac{1}{2 \, y_1 \, y_2 \mathcal{P}_{(1,2),(1,p),(2,p)} \mathcal{P}_{(2,1),(1,p),(2,p)}}\, ,
\end{equation}
where again the $\to$ stands for taking the residue in $\tilde{c}$. One can now easily check that the three terms obtained above all correspond to terms entering in the contribution of the bubble diagram to the $2$-point correlator given in \eqref{eq:bubble_corr}.

\bibliographystyle{apsrev4-1.bst}
\bibliography{Refs.bib}

\end{document}